\documentclass[11pt]{JHEP3}
\usepackage{latexsym}
\usepackage{graphics}
\usepackage{amsmath,amsfonts}

\def\NO{\nonumber}

\newcommand{\be}{\begin{equation}}
\newcommand{\ee}{\end{equation}}

\def\bea{\begin{eqnarray}}
\def\eea{\end{eqnarray}}

\def\beqx{\begin{displaymath}}
\def\eeqx{\end{displaymath}}

\newcommand{\bmat}{\left(\begin{array}}
\newcommand{\emat}{\end{array}\right)}

\def\a{\alpha}
\def\b{\beta}

\def\d{\delta}
\def\e{\epsilon}
\def\f{\phi}
\def\g{\gamma}
\def\h{\eta}

\def\k{\kappa}
\def\l{\lambda}
\def\m{\mu}
\def\n{\nu}

    \def\om{\omega}
\def\p{\pi}

    \def\th{\theta}
\def\r{\rho}
\def\s{\sigma}

\def\x{\xi}

\def\D{\Delta}

\def\G{\Gamma}

\def\L{\Lambda}

    \def\Om{\Omega}
\def\P{\Pi}

\def\S{\Sigma}

\def\ve{\varepsilon}

\def\vf{\varphi}

\def\cg{{\cal G}}

\def\ci{{\cal I}}

\def\cl{{\cal L}}
\def\cm{{\cal M}}
\def\cn{{\cal N}}
\def\co{{\cal O}}
\def\cp{{\cal P}}

\def\bo{{\raise-.3ex\hbox{\large$\Box$}}}               
\def\pa{\partial}                                       
\def\face{{\raise.2ex\hbox{$\displaystyle \bigodot$}\mskip-2.2mu \llap {$\ddot
        \smile$}}}                                   
\def\>{\rangle}                                      
\def\<{\langle}                                      

\newcommand{\sub}[1]{\phantom{}_{(#1)}\phantom{}}    
\def\lbar#1{\ensuremath{\overline{#1}}}              
\def\leftrightarrowfill{$\mathsurround=0pt \mathord\leftarrow \mkern-6mu
        \cleaders\hbox{$\mkern-2mu \mathord- \mkern-2mu$}\hfill
        \mkern-6mu \mathord\rightarrow$}        
\def\dvec#1{\vbox{\ialign{##\crcr
        \leftrightarrowfill\crcr\noalign{\kern-1pt\nointerlineskip}
        $\hfil\displaystyle{#1}\hfil$\crcr}}}           
\def\tr{{\rm tr \,}}                                    
\def\Tr{{\rm Tr \,}}                                    
\def\diag{{\rm diag \,}}                                

\def\-{\hphantom{-}}

\title{Non-Supersymmetric Membrane Flows from Fake Supergravity and
Multi-Trace Deformations}

\author{Ioannis Papadimitriou \\
DESY Theory Group,\\ Notkestrasse 85,\\ D-22603 Hamburg, Germany \\ \\
and\\ \\
Center for Mathematical Physics, \\
Bundesstrasse 55,\\
D-20146 Hamburg, Germany.\\ \\
\email{ioannis.papadimitriou@desy.de}}

\abstract{We use fake supergravity as a solution generating technique
to obtain a continuum of non-supersymmetric asymptotically $AdS_4\times S^7$
domain wall solutions of eleven-dimensional supergravity with non-trivial
scalars in the $SL(8,\mathbb{R})/SO(8)$ coset. These solutions are
continuously connected to the supersymmetric domain walls describing a uniform
sector of the Coulomb branch of the $M2$-brane theory. We also provide a
general argument that under certain conditions identifies the fake 
superpotential with the exact large-$N$ quantum effective potential of the 
dual theory, describing a marginal multi-trace deformation. This identification
strongly motivates further study of fake supergravity as a solution generating 
method and it allows us to interpret our non-supersymmetric solutions as a 
family of marginal triple-trace deformations of the Coulomb branch that 
completely break supersymmetry and to calculate the exact large-$N$ anomalous 
dimensions of the operators involved. The holographic one- and two-point 
functions for these solutions are also computed.}

\preprint{DESY-06-083\\
ZMP-HH/06-09\\
hep-th/0606038}

\begin{document}

\tableofcontents
\addtocontents{toc}{\protect\setcounter{tocdepth}{2}}

\section{Introduction and summary of results}
\setcounter{equation}{0}

The study of domain wall solutions of various supergravity theories
has been strongly motivated in recent years by the role these play
in a variety of physical contexts, from the AdS/CFT correspondence,
where they describe an RG flow of the conformal field theory residing on
the conformal boundary of AdS, to `Brane World' scenarios and cosmological
models (see \cite{Cvetic:1996vr} for an extensive review of domain walls
of $N=1$ supergravity in four dimensions). Although, when they arise as 
solutions to a particular supergravity theory, such domain walls are often 
supersymmetric, this need not be the case. Indeed, many non-supersymmetric 
gravitational theories admit domain wall solutions as well. In this paper, 
however, we will emphasize the fact that true supergravity theories also 
admit non-supersymmetric domain wall solutions, which can be physically 
important.

We will focus on domain walls preserving Poincar\'e invariance in $d=D-1$
dimensions, where $D$ is the spacetime dimension where the given
gravitational theory lives. Such domain walls take the form\footnote{More
general domain walls with a different isometry do exist, as is discussed e.g.
in \cite{Freedman:2003ax}, but we will not discuss them here.}
\be\label{poincare}
ds_D^2=dr^2+e^{2A(r)}\h_{ij}dx^idx^j,\quad \f^I=\f^I(r),
\ee
where $\h=\diag(-1,1,\cdots,1)$ is the Minkowski metric in $d$ dimensions.
Since only the metric and a number of scalar fields are involved in these
solutions, they can generically be described by an effective gravitational
theory with an action of the form
\be\label{fake_action}
S=\int_\cm d^{D}x\sqrt{-g}\left(\frac{1}{2\k^2_D}R-\frac12\cg_{IJ}(\f)
g^{\m\n}\pa_\m\f^I\pa_\n\f^J-V(\f)\right),
\ee
where $\k_D^2=8\p G_{D}$ is the effective gravitational constant and
$\cg_{IJ}$ is a generic (Riemannian) metric on the scalar manifold. Such
theories arise naturally as consistent truncations of various gauged
supergravities, in which case the scalar potential is generated by
the non-trivial gauging of (some of) the isometries of the scalar manifold.
Generically, however, this effective description will only be valid
locally in the moduli space of a given supergravity theory \cite{Celi:2004st}.
Here we are interested in the application of domain walls to the AdS/CFT
correspondence and so we assume that the metric (\ref{poincare}) is
asymptotically AdS, which is equivalent to the statement that $A(r)\sim r$
as $r\to\infty$. This in turn implies that the scalar potential $V(\f)$ has
at least one stable fixed point at $\f^I=\f^I_*$ such that $V(\f_*)<0$. By a
reparameterization of the scalar manifold we can set $\f^I_*=0$. If
this potential arises from some gauged supergravity, this fixed point
corresponds to the maximally symmetric $AdS_{D}$ vacuum.

The equations of motion following from the action (\ref{fake_action}) are
Einstein's equations
\be
R_{\m\n}-\frac12 Rg_{\m\n}=\k_D^2T_{\m\n},
\ee
with the stress tensor given by
\be
T_{\m\n}=\cg_{IJ}(\f)\pa_\m\f^I\pa_\n\f^J-g_{\m\n}\left(\frac12\cg_{IJ}(\f)
g^{\r\s}\pa_\r\f^I\pa_\s\f^J+V(\f)\right),
\ee
and
\be
\nabla^\m\left(\cg_{IJ}(\f)\pa_\m\f^J\right)-\frac12\frac{\pa \cg_{LM}}{\pa
\f^I}g^{\m\n}\pa_\m\f^L\pa_\n\f^M-\frac{\pa V}{\pa\f^I}=0.
\ee
Substituting the domain wall ansatz (\ref{poincare}) into the equations of
motion one obtains the following equations for the warp factor $A(r)$ and the
scalar fields $\f^I(r)$:
\bea
&&\dot{A}^2-\frac{\k^2}{d(d-1)}\left(\cg_{IJ}(\f)\dot{\f}^I\dot{\f}^J-2V(\f)
\right)=0,\NO\\
&&\ddot{A}+d\dot{A}^2+\frac{2\k^2}{d-1}V(\f)=0,\NO\\\label{2nd-order-eqs}
&&\cg_{IJ}(\f)\ddot{\f}^J+\frac{\pa \cg_{IJ}}{\pa\f^K}\dot{\f}^K\dot{\f}^J
-\frac12\frac{\pa \cg_{LM}}{\pa \f^I}\dot{\f}^L\dot{\f}^M+d\dot{A}\cg_{IJ}(\f)
\dot{\f}^J-\frac{\pa V}{\pa\f^I}=0,
\eea
where the dot denotes the derivative with respect to the radial coordinate $r$.

It is important to distinguish between two types of solutions of these
second order equations. Following \cite{Skenderis:1999mm} we will call
a `BPS domain wall' any domain wall of the form (\ref{poincare}) which
satisfies the first order equations
\bea\label{flow_eqs}
&&\dot{A}=-\frac{\k^2}{d-1}W(\f),\NO\\
&&\dot{\f}^I=\cg^{IJ}(\f)\frac{\pa W}{\pa \f^J},
\eea
for some function $W(\f)$ of the scalar fields such that the scalar
potential can be expressed as
\be\label{potential}
V(\f)=\frac12\left(\cg^{IJ}(\f)\frac{\pa W}{\pa\f^I}\frac{\pa W}{\pa\f^J}-
\frac{d\k^2}{d-1}W^2\right).
\ee
Note that the first order equations (\ref{flow_eqs}) together with
(\ref{potential}) ensure that the second order equations (\ref{2nd-order-eqs})
are automatically satisfied. Given the expression (\ref{potential}) for the
scalar potential in terms of the function $W$, the first order equations
(\ref{flow_eqs}) can be derived \'a la Bogomol'nyi by extremizing the energy
functional $E[A,\f]$ that has (\ref{2nd-order-eqs}) as its Euler-Lagrange
equations \cite{Skenderis:1999mm, Bakas:1999fa}. In the context of gauged
supergravity, a function $W(\f)$ satisfying (\ref{potential}) arises naturally
as the superpotential, $W_o(\f)$, which enters the gravitino and dilatino
variations
\bea\label{susy}
&&\d\psi_\m=D_\m\ve-\frac{\k^2}{2(d-1)}W_o(\f)\g_\m\ve,\\
&&\d\chi^I=\left(\g^\m\pa_\m\f^I+\cg^{IJ}(\f)\frac{\pa W_o}{\pa\f^J}\right)\ve.
\eea
It follows that the domain walls defined by the superpotential $W_o(\f)$
are supersymmetric solutions of the particular gauged supergravity.
Crucially, however, equation (\ref{potential}) does {\em not} define
the function $W(\f)$ uniquely and hence there may generically exist other
functions $W(\f)$ satisfying (\ref{potential}) in addition to
$W_o(\f)$.\footnote{Note, however, that not every function $W(\f)$ that
satisfies (\ref{potential}) is acceptable, since it will not generically
correspond to an asymptotically AdS domain wall. We will discuss in detail
the conditions $W(\f)$ must satisfy below. See also
\cite{Freedman:2003ax, Skenderis:1999mm}.} This has been termed {\em fake
supergravity} and in this context any function $W(\f)$ that solves
(\ref{potential}) is called a {\em fake superpotential}
\cite{Freedman:2003ax,Celi:2004st,Zagermann:2004ac,Skenderis:2006jq}.
In \cite{Skenderis:2006jq} it was shown that any BPS Poincar\'e domain wall of
the form (\ref{poincare}), defined by a function $W(\f)$ which is not
necessarily the true superpotential of a given gauged supergravity, is
`supersymmetric' in the sense that one can always find Killing spinors, at
least locally. In \cite{Skenderis:2006jq} this was considered as an
indication that {\em any} function $W(\f)$ that solves (\ref{potential})
(and possibly subject to suitable boundary conditions) may be the true
superpotential of {\em some} supergravity theory, even though there is no
systematic way to find which is the relevant, known or unknown, theory
\cite{kostas}. Despite the elegance of this statement, it is difficult in
practice to confirm or refute it. We will adopt a rather different point of
view here, however. Namely, we will confine ourselves to a {\em particular}
gauged supergravity, with a certain superpotential $W_o(\f)$. Clearly, any
BPS domain wall defined by a solution $W(\f)\neq W_o(\f)$ of (\ref{potential})
is {\em not} supersymmetric in this context. We will nevertheless continue to
call such solutions `BPS' since they satisfy the first order equations
(\ref{flow_eqs}). They are still special solutions because they allow for the
definition of {\em fake Killing spinors} via (\ref{susy}) with $W_o(\f)$
replaced by the fake superpotential $W(\f)$
\cite{Freedman:2003ax}.\footnote{Note that our fake superpotential differs
by a factor of $-\frac{2(d-1)}{\k^2}$ relative
to the fake superpotential defined in \cite{Freedman:2003ax}. The
supercovariant derivative is uniquely determined, however, by the requirement
that it reduces to that of pure AdS, namely $(D_\m+\frac{1}{2l}\g_\m)\ve$,
when the scalar fields vanish.} The existence of fake Killing spinors implies,
in particular, non-perturbative gravitational stability, at least in the
absence of naked singularities \cite{Townsend:1984iu, Skenderis:1999mm,
Freedman:2003ax}.

If the scalar potential cannot be written in the form (\ref{potential}),
however, there can still exist domain wall solutions of the form
(\ref{poincare}) that solve the second order equations (\ref{2nd-order-eqs}).
We will refer to such solutions as `non-BPS domain walls'.\footnote{Note
though the analysis of \cite{Skenderis:2006jq}, which suggests that
any potential that admits domain wall solutions can be written in the
form (\ref{potential}) and so there are no `non-BPS' domain walls.} We will not
consider further such domain walls here since we are interested in scalar
potentials that arise from gauged supergravities, and such potentials
are guaranteed to be expressible in the form (\ref{potential}) since
this is at least possible using the true superpotential
$W_o(\f)$.\footnote{Generically the superpotential $W_o(\f)$ will be a
matrix, however, instead of a scalar quantity. See e.g.
\cite{Freedman:2003ax}.}

Although one often views fake supergravity as an effective
subsector of some gauged supergravity, by identifying both the scalar potential
and the fake superpotential of fake supergravity with the true potential and
superpotential respectively of the gauged supergravity
\cite{Skenderis:1999mm,Celi:2004st,Zagermann:2004ac}, we will instead treat
fake supergravity as a powerful solution generating technique for
{\em non-supersymmetric} solutions of a given gauged supergravity.
In particular, we will treat (\ref{potential}) as a first order non-linear
differential equation for the fake superpotential $W(\f)$
\cite{DeWolfe:1999cp, Gubser:2000nd, Freedman:2003ax,Papadimitriou:2004ap}
(see also \cite{Campos:2000yu} where a very similar perspective is adopted).
For scalar potentials arising from some gauged supergravity this equation
admits at least one solution, namely the true superpotential of the theory.
Our aim here will be to determine {\em all} solutions of (\ref{potential})
satisfying appropriate boundary conditions. Each solution $W(\f)\neq W_o(\f)$
defines a non-supersymmetric domain wall solution of the given gauged
supergravity, and therefore describes a non-supersymmetric RG flow
of the dual field theory.

The paper is organized as follows. In the next section we will discuss
a common subsector of gauged maximal supergravities in dimensions $D=4,5,7$
with the scalar fields parameterizing an $SL(N,\mathbb{R})/SO(N)$ coset,
where $N=8,6,5$ respectively. The complete non-linear ansatz for
uplifting any solution of this subsector to eleven-dimensional or Type IIB
supergravity is known \cite{Cvetic:1999xx,Cvetic:2000eb}, and all
supersymmetric Poincar\'e domain walls, describing a uniform subsector
of the Coulomb branch of respectively the $M2$-, $D3$-, or $M5$-brane theory
have been constructed \cite{Kraus:1998hv, Freedman:1999gk, Bakas:1999ax,
Cvetic:1999xx, Bakas:1999fa}. In Section \ref{domain-walls} we solve
equation (\ref{potential}) with the scalar potential of gauged
supergravity as a differential equation for the fake superpotential $W(\f)$,
subject to suitable boundary conditions. We show that analytic 
non-supersymmetric
solutions exist only in dimension $D=4$, while the superpotential $W_o$
is the only analytic solution of (\ref{potential}) for $D=5,7$.
In Section \ref{D=4-domain-walls} we systematically discuss how to
obtain these non-supersymmetric solutions in four dimensions in closed
form by consistently reducing the number of scalar fields, and we
solve (\ref{potential}) exactly, obtaining a family of exact
fake superpotentials, for a special case involving a single scalar field.
We then uplift this solution to eleven dimensions in Section \ref{uplift}
using the ansatz discussed in Section  \ref{gauged-sugra} and, noting
that the MTZ black hole \cite{Martinez:2004nb} in four dimensions is
interestingly a solution of exactly the same action as our exact domain
wall,  we also give the eleven-dimensional black hole solution (given
explicitly in Appendix \ref{MTZ}). The holographic one- and two-point
functions for the non-supersymmetric domain walls are then computed
respectively in Sections \ref{1-pt-fns} and \ref{2pt-functions}. Finally,
in Section \ref{eff-pot} we show that under certain circumstances the fake 
superpotential $W(\f)$ that solves (\ref{potential}) and corresponds to an 
asymptotically AdS domain wall, defines a marginal multi-trace
deformation of the dual field theory. This means that solving
equation (\ref{potential}) as a differential equation for the fake
superpotential not only is interesting as a method for finding exact
non-supersymmetric supergravity solutions, but also, these solutions can
often be interpreted as the exact holographic duals of a marginal 
multi-trace deformation of the boundary theory. Applying this
observation to the non-supersymmetric domain walls we have constructed
leads to the conclusion that they correspond to a continuous family
of marginal triple-trace deformations of the Coulomb branch of the $M2$-brane
theory. A number of technical results are collected in the appendices.

\section{The $SL(N,\mathbb{R})/SO(N)$ sector of gauged maximal 
supergravity and its higher-dimensional origin}
\label{gauged-sugra}

The scalar manifold of $D$-dimensional maximal supergravity is
the coset $E_{11-D(11-D)}/K$, where $E_{n(n)}$ is the maximally non-compact
form of the exceptional Lie group $E_n$ and $K$ is its maximal compact
subgroup.\footnote{For $n<6$ the following identifications are made
$E_5\cong D_5,\; E_4\cong A_4,\; E_3\cong A_2\times
A_1,\; E_2\cong A_1\times \mathbb{R}$ and $E_1\cong\mathbb{R}$.}
Following \cite{Bakas:1999ax, Cvetic:1999xx, Bakas:1999fa}, we specialize
to an $SL(N,\mathbb{R})$ subgroup of $E_{11-D}$, where
$N=4(D-2)/(D-3)$, and consider the $\frac12 N(N+1)-1$ scalars of the
coset $SL(N,\mathbb{R})/SO(N)$. This scalar sector is common to all maximal
supergravities in any dimension. In particular, for $D=7$ we consider
$SL(5,\mathbb{R})\cong E_4$, for $D=5$
$SL(6,\mathbb{R})\subset E_6$ and for $D=4$ $SL(8,\mathbb{R})\subset E_7$.

The Lagrangian density describing this sector of the gauged version of maximal
supergravity in $D$ dimensions, where the $SO(N)$ symmetry is gauged, is
\be\label{lagrangian1}
e^{-1}\cl_D=\frac{1}{2\k_D^2}R+\frac{1}{8\k_D^2}\tr
\left(\pa_\m\cm\pa^\m\cm^{-1}\right)-V,
\ee
where $\cm=S^TS$ is a symmetric $N\times N$ matrix, with $S$ in the
fundamental representation of $SL(N,\mathbb{R})$, and the potential $V$
takes the form\footnote{Note that the $AdS_D$ radius, $l_D$, is related to
the coupling $g$ in \cite{Cvetic:1999xx,Cvetic:2000eb} by $l_D=(D-3)/2g$.}
\be\label{potential1}
V=-\frac{(D-3)^2}{16\k_D^2l_D^2}\left[(\tr\cm)^2-2\tr(\cm^2)\right].
\ee
In these expressions the trace is taken in the fundamental of
$SL(N,\mathbb{R})$. Using an $SO(N)$ rotation, the matrix $\cm$ can
be diagonalized so that
\be
\cm={\rm diag}(X_1,\ldots,X_N),
\ee
where the $N$ scalars $X_i$ satisfy the constraint
\be\label{constraint}
\det\cm=\prod_{i=1}^NX_i=1.
\ee
It might be useful to note that in terms of the non-trivial scalars that
we have kept at this point, the symmetric tensor $T_{ij}$ parameterizing
the full scalar manifold of the maximal supergravity takes the form
$T_{ij}=X_i\d_{ij}$. The $N$ constrained scalars $X_i$ can be parameterized
by $N-1$ independent scalar fields, $\vf^I$, $I=1,\ldots,N-1$, as
\be
X_i=e^{-\frac12\vec{b}_i\cdot\vec{\vf}},
\ee
where the $N$ vectors $\vec{b}_i$ are (up to a factor of 2) the
weight vectors of the fundamental representation of $SL(N,\mathbb{R})$
and they satisfy
\be\label{weights}
\vec{b}_i\cdot\vec{b}_j=8\d_{ij}-\frac8N,\quad \sum_{i=1}^N\vec{b}_i=0,\
\quad \sum_{i=1}^Nb_{iI}b_{iJ}=8\d_{IJ}.
\ee

After diagonalizing the matrix $\cm$ and dropping the kinetic
terms for the original off-diagonal scalars which decouple, the Lagrangian
(\ref{lagrangian1}) becomes
\be\label{lagrangian2}
e^{-1}\cl_D=\frac{1}{2\k_D^2}R-\frac{1}{4\k_D^2}\sum_{I=1}^{N-1}\pa_\m\vf^I
\pa^\m\vf^I-V,
\ee
where the potential is now given by
\be\label{potential2}
V=-\frac{(D-3)^2}{16\k_D^2l_D^2}\left((\sum_{i=1}^{N}X_i)^2-
2\sum_{i=1}^{N}X_i^2\right).
\ee
This Lagrangian, which is a special case of (\ref{fake_action}), falls into
the framework of fake supergravity described in the previous section. In
order to make contact with our notation in the previous section we
also define the rescaled scalars
\be
\f^I\equiv \frac{1}{\sqrt{2}\k_D}\vf^I,
\ee
which have a canonically normalized kinetic term.

The equations of motion for this gravity-scalar system can be written as
\bea\label{eqs_motion}
&&R_{\m\n}=\frac{1}{4}\sum_{i=1}^NX_i^{-2}\pa_\m X_i\pa_\n X_i+
\frac{2\k_D^2}{D-2}Vg_{\m\n},
\NO\\
&&\square\log X_i=\frac{(D-3)^2}{2l_D^2}\left(2X_i^2-X_i\sum_{j=1}^NX_j-
\frac2N\sum_{j=1}^NX_j^2+\frac1N(\sum_{j=1}^NX_j)^2\right).
\eea
The second of these equations can be derived by starting from the equation
of motion for the independent scalar fields $\vf^I$,
\be
\square\vf^I=\frac{(D-3)^2}{8l_D^2}\sum_{i=1}^Nb_{iI}X_i(\sum_{j=1}^NX_j-2X_i),
\ee
noticing that the last equation in (\ref{weights}) implies that
\be
\square\vf^I=-\frac14\sum_{i=1}^Nb_{iI}\log X_i,
\ee
and adding a term to ensure that the sum over $i$ is zero, in agreement
with the constraint (\ref{constraint}).

The gravity-scalar theory we have just discussed was obtained as a consistent
truncation of gauged maximal supergravity in $D$ dimensions. However,
the maximal gauged supergravities in $D=4$ and $D=7$ are known to
arise themselves as consistent truncations to the massless fields of the
Kaluza-Klein compactification of eleven-dimensional supergravity on $S^7$ and
$S^4$ respectively \cite{deWit:1986iy,Nastase:1999cb,Nastase:1999kf}.
Moreover, the gauged maximal supergravity in $D=5$ is also believed to arise
as an $S^5$ reduction of Type IIB supergravity, although a full proof is
still lacking. It is therefore expected that the above gravity-scalar
theory should also be obtainable directly as a consistent truncation
of eleven-dimensional or Type IIB supergravity. Indeed, the full non-linear
ansatz for this reduction, valid for any $D$, was given in
\cite{Cvetic:1999xx} and it was later proved in \cite{Cvetic:2000eb} that
this is a consistent truncation of the higher-dimensional theory, that is,
the equations of motion of the higher dimensional theory with the
ansatz (\ref{ansatz}) are satisfied if and only if the equations of
motion for the gravity-scalar system (\ref{eqs_motion}) are satisfied in
$D$ dimensions.

The reduction ansatz given in \cite{Cvetic:1999xx} is
\bea\label{ansatz}
d\hat{s}^2 & = & \D^{\frac{2}{D-1}}ds_D^2+\frac{4l_D^2}{(D-3)^2}
\D^{-\left(\frac{D-3}{D-1}\right)}\sum_{i=1}^NX_i^{-1}d\m_i^2,\\
\hat{F}\sub{D} & = & \frac{(D-3)}{2l_D}\sum_{i=1}^N(2X_i^2\m_i^2-\D X_i)
\e\sub{D}-\frac{l_D}{(D-3)}\sum_{i=1}^NX_i^{-1}\ast_D dX_i\wedge d(\m_i^2),
\NO
\eea
where
\be
\D=\sum_{i=1}^NX_i\m_i^2,
\ee
and $\m_i$ stand for a set of $N$ direction cosines satisfying
\be
\sum_{i=1}^N\m_i^2=1.
\ee
Moreover, $\e\sub{D}$ denotes the volume form of the metric $ds_D^2$, while
the field strength $\hat{F}\sub{D}$ is identified with the M-theory four-form
for $D=4$, its Hodge dual for $D=7$, and with the self-dual five-form of
IIB supergravity for $D=5$.

\section{All asymptotically AdS Poincar\'e domain walls of the
$SL(N,\mathbb{R})/SO(N)$ sector}
\label{domain-walls}

From the discussion of the previous section we know that any
solution of the equations of motion (\ref{eqs_motion}) in $D$ dimensions
can be uplifted to solutions of either eleven-dimensional or Type IIB
supergravity. In particular, any Poincar\'e domain wall of the
form (\ref{poincare}) corresponds to a solution of the
higher-dimensional theory. Indeed, all {\em supersymmetric}
asymptotically $AdS_D$ domain walls in $D=4,5,7$ have been
constructed \cite{Bakas:1999ax, Cvetic:1999xx, Bakas:1999fa}.\footnote{The
case $D=6$, corresponding to an $S^4$ reduction of massive Type IIA
was also considered in \cite{Cvetic:1999xx}.} These domain walls solve
the first order equations (\ref{flow_eqs}) with the true superpotential
of the $SL(N,\mathbb{R})/SO(N)$ sector of gauged maximal supergravity,
which takes the form
\be\label{superpotential_0}
W_o=-\frac{(D-3)}{4\k_D^2l_D}\sum_{i=1}^NX_i.
\ee
It can be easily verified that this superpotential solves (\ref{potential})
with the scalar potential (\ref{potential2}). The uplifted solutions
are asymptotically $AdS_4\times S^7$, $AdS_5\times S^5$ or $AdS_7\times S^4$
and correspond to continuous distributions of parallel $M2$-, $D3$- or
$M5$-branes respectively. Generically, they contain naked null singularities,
corresponding to the location of the continuous brane distribution.

It was argued in \cite{Bakas:1999ax, Cvetic:1999xx, Bakas:1999fa},
following \cite{Kraus:1998hv,Freedman:1999gk}, that these
supersymmetric solutions describe the RG flow of the dual CFTs
due to the VEV of the scalar operators dual to the $SL(N,\mathbb{R})/SO(N)$
scalars, with the VEVs defined by the brane distribution. Although in $D=5,7$
this interpretation is unique due to the unambiguous identification of the
$SL(N,\mathbb{R})/SO(N)$ scalars as dual to operators of dimension 2 and
4 respectively, in $D=4$ there is an ambiguity in the dimension of the
operators dual to the $SL(8,\mathbb{R})/SO(8)$ scalars. This because,
as we will explain in detail below, the scalar potential (\ref{potential2})
implies that the $SL(N,\mathbb{R})/SO(N)$ scalars, except for $D=5$ in which
case the mass saturates the Breitenlohner-Freedman (BF) bound
\cite{Breitenlohner:1982jf}, have a mass that allows their association with
operators of {\em two} possible dimensions instead of one
\cite{Klebanov:1999tb}. For $D=7$, however, this ambiguity is removed by
symmetry. Namely, only the scalars of dimension 4 appear in the massless
$\cn=2$ supermultiplet. For $D=4$ the 35 dimension 1 scalars and the 35 
dimension 2 scalars both appear in the massless $\cn=8$ supermultiplet 
on an equal footing. Therefore, although the interpretation
of these solutions in terms of VEVs for the dual operators remains correct,
for $D=4$ there is a second possible interpretation in terms of deformations
of the CFT Lagrangian. We will analyze this issue carefully below, when we
compute in complete generality the VEVs of the possible dual operators.

In this section, however, we will try to systematically find all
asymptotically AdS Poincar\'e domain wall solutions of the
$SL(N,\mathbb{R})/SO(N)$ sector, and in particular, all
{\em non-supersymmetric} ones. In other words, we will
determine the most general fake superpotential $W(\f)$ satisfying
(\ref{potential}) with the scalar potential given by
(\ref{potential2}).\footnote{As we have mentioned already, the fake
superpotential can in general be matrix valued, but we will only
analyze the case of a scalar fake superpotential here.} We will find
that for $D=5$ and $D=7$, there are no analytic non-supersymmetric 
asymptotically AdS Poincar\'e domain walls. For $D=4$, however, we will show 
that there exists a continuum of analytic non-supersymmetric domain walls, 
continuously connected to the supersymmetric ones, as well as, a number of 
isolated non-supersymmetric domain walls.

Equation (\ref{potential}) is a first order non-linear PDE in $N-1$ variables
for the fake superpotential $W(\f)$, and as such, solving this equation in
full generality for $W(\f)$ seems a rather formidable task. However, not
all solutions (\ref{potential}) are physically admissible, if one is
interested in asymptotically AdS domain walls. In particular, the requirement
that the domain walls defined by $W(\f)$ via (\ref{flow_eqs}) asymptote
to AdS space as $\f^I\to 0$ implies that
\be\label{W_zero}
W(0)=-\frac{(d-1)}{\k_D^2 l_D}.
\ee
In addition we will assume that $W(\f)$ admits a Taylor
expansion around the maximally symmetric fixed point of $V(\f)$ corresponding
to $\f^I=0$, namely
\be\label{W_Taylor}
W(\f)=\sum_{n=0}^\infty W\sub{n}_{I_1\ldots I_n}\f^{I_1}\cdots\f^{I_n},
\ee
where all coefficients are completely symmetric in their indices, and
$W\sub{0}=W(0)$ is given by (\ref{W_zero}). Within this framework,
analyzing (\ref{potential}) in full generality is now tractable.

We start by Taylor expanding the scalar potential (\ref{potential2}) around
$\f^I=0$. We find
\bea\label{V_Taylor}
V & = & \sum_{n=0}^\infty V\sub{n}_{I_1\ldots I_n}\f^{I_1}\cdots\f^{I_n}\\
& = & -\frac{d(d-1)}{2\k_D^2l_D^2}+\frac12  m_I^2 \f^I\f^I
-\frac{(d-2)(d-3)\sqrt{2}\k_D}{48l_D^2}\sum_{i=1}^Nb_{iI}b_{iJ}b_{iK}
\f^I\f^J\f^K+\co(\f^4),\NO
\eea
where $m_I^2l_D^2=\D_I(\D_I-d)=2(2-d)$. Inserting the expansions
for $V(\f)$ and $W(\f)$ in (\ref{potential}) and matching powers one
obtains the following recursion relations for the coefficients $W\sub{n}$:
\bea\label{recursion}
\sum_{m=0}^n\left[\rule[.2cm]{0cm}{.3cm}(m+1)(n-m+1)W\sub{m+1}_{(I_1\ldots 
I_{m}J}
W\sub{n-m+1}^{J}\phantom{}_{I_{m+1}\ldots I_{n})_s}\right.\NO\\\left.
-\frac{d\k_D^2}{d-1}
W\sub{m}_{(I_1\ldots I_{m}}W\sub{n-m}_{I_{m+1}\ldots I_{n})_s}\right]
=2V\sub{n},
\eea
where $(\ldots)_s$ denotes symmetrization with weight 1. For $n=0$ the
recursion relations give
\be
W\sub{1}_JW\sub{1}^J-\frac{d\k_D^2}{d-1}W\sub{0}^2=2V\sub{0}.
\ee
Using the values for $W\sub{0}$ and $V\sub{0}$ from (\ref{W_zero}) and
(\ref{V_Taylor}) we deduce that
\be
W\sub{1}_I=0.
\ee
Since $V\sub{1}_I=0$, which is guaranteed on general grounds by the
requirement that AdS is a fixed point of the scalar potential, the
equation for $n=1$, which reads
\be
\left(2W\sub{2}_{IJ}+\frac{d}{l_D}\d_{IJ}\right)W\sub{1}^J=
V\sub{1}_I,
\ee
is automatically satisfied. Using the fact that $W\sub{1}_I=0$, the next two
equations now take the form
\bea
n=2: & &\label{W2_eq}
\left(2W\sub{2}_{IJ}+\frac{d}{l_D}\d_{IJ}\right)W\sub{2}^J
\phantom{}_K=V\sub{2}_{IK},\\\NO\\
n=3: & &\label{W3_eq}
\left(6W\sub{2}_{IJ}+\frac{d}{l_D}\d_{IJ}\right)W\sub{3}^J
\phantom{}_{KL}=V\sub{3}_{IKL},
\eea
while for higher $n$ the recursion relations give
\be
\left(2n W\sub{2}_{IJ}+\frac{d}{l_D}\d_{IJ}\right)W\sub{n}^J
\phantom{}_{K_1\ldots K_{n-1}}+\ldots=V\sub{n}_{IK_1\ldots K_{n-1}},
\ee
where the dots stand for terms involving the coefficients $W\sub{m}$ with
$m<n$. It follows that, given the symmetric  matrix $W\sub{2}_{IJ}$, the
recursion relations uniquely determine all higher coefficients of $W(\f)$,
{\em unless} the matrix
\be\label{degeneracy_matrix}
\left(2n W\sub{2}_{IJ}+\frac{d}{l_D}\d_{IJ}\right),
\ee
has some zero eigenvalues for some $n>2$. To address the question if and
when this can happen we first have to solve equation (\ref{W2_eq}) which
determines $W\sub{2}_{IJ}$.

From (\ref{V_Taylor}) we see that $V\sub{2}_{IJ}=-\frac{(d-2)}{l_D^2}\d_{IJ}$.
Since $W\sub{2}_{IJ}$ is a symmetric matrix it can be diagonalized by an
orthogonal matrix $R_{IJ}$. Such a (rigid) rotation in the space of the
$N-1$ independent scalars would leave the form of the potential invariant
since it simply rotates the weights $b_i$, while preserving the
relations (\ref{weights}). Hence, we can take $W\sub{2}_{IJ}$ to be diagonal:
$W\sub{2}_{IJ}=w_I\d_{IJ}$. Equation (\ref{W2_eq}) then reduces to $N-1$
decoupled equations for the diagonal components, $w_I$, of $W\sub{2}_{IJ}$,
namely
\be
(2w_I+d/l_D)w_I+(d-2)/l_D^2=0, \quad I=1,\ldots,N-1,
\ee
where there is no summation implied in this equation. The roots of this
equation are $w_I=w_\pm$, where $w_+=-1/l_D$, $w_-=-(d-2)/2l_D$, and hence,
for $d\neq 4$, there are $2^{N-1}$ independent solutions
$W\sub{2}=\diag(w_\pm,\ldots,w_\pm)$,
corresponding to the possible distributions of $w_\pm$ along the diagonal.
For $d=4$ however, $w_+$ and $w_-$ coincide and there is a unique
solution for $W\sub{2}$.
It follows that the matrix (\ref{degeneracy_matrix}) is diagonal with
diagonal values $2n w_\pm+d/l_D$. Now, $2n w_++d/l_D=(d-2n)/l_D$ can vanish if
$d$ is even, i.e. $d=4,6$ since we are interested in the cases $d=3,4,6$.
Similarly, $2n w_-+d/l_D=(d-n(d-2))/l_D$ can vanish if $d/(d-2)$ is integer,
i.e. for $d=3,4$. However, in either case, $d=4$ requires $n=2$, which is
excluded since we have already determined $W\sub{2}$. It follows that for
$d=4$, the true superpotential $W_o$ given in (\ref{superpotential_0}) is
the {\em unique} (physical) solution of (\ref{potential}). For $d=3,6$,
however, we have seen that there are $2^{N-1}$ choices for $W\sub{2}$ and
for each of them there is possibly some freedom in the value of $W\sub{3}$
due to the vanishing of some of the eigenvalues of the matrix
(\ref{degeneracy_matrix}), but all higher coefficients in $W(\f)$ are
completely determined once a choice for $W\sub{2}$ and $W\sub{3}$ has been
made. Equation (\ref{W3_eq}) however imposes further constraints. Noticing
from (\ref{V_Taylor}) that $V\sub{3}$ vanishes for $d=3$ but not for $d> 3$,
equation (\ref{W3_eq}) implies that for $d=6$, the matrix
(\ref{degeneracy_matrix}) must have no zero eigenvalues and therefore both
$W\sub{2}=\diag(w_-,\ldots,w_-)$ and $W\sub{3}$ are uniquely determined. So,
as for $d=4$, $W_o$ in (\ref{superpotential_0}) is the unique solution of
(\ref{potential}).  For $d=3$, however, $V\sub{3}$ vanishes identically and
so either the matrix (\ref{degeneracy_matrix}) vanishes identically or
$W\sub{3}$ vanishes identically. In the first case
$W\sub{2}=\diag(w_-,\ldots,w_-)$ and $W\sub{3}$ is arbitrary, while in the
second case $W\sub{3}=0$ and $W\sub{2}$ can be any of the $2^7$ possible
diagonal matrices. We conclude that $d=3$ is the only case which allows
additional Poincar\'e domain wall solutions beyond the supersymmetric ones
corresponding to the superpotential (\ref{superpotential_0}). We will
now examine these solutions more closely and construct explicitly
as many of these as possible.

\section{The non-supersymmetric Poincar\'e domain walls in $D=4$}
\label{D=4-domain-walls}

As we have just shown, only in four dimensions ($d=3$) are there physically
acceptable solutions $W(\f)$ to (\ref{potential}), in addition to the
supersymmetric solution (\ref{superpotential_0}). These solutions
fall into two general classes. The first case is when $W\sub{3}=0$ and
$W\sub{2}=\diag(w_\pm,\ldots, w_\pm)$, where all signs are chosen
independently. There are therefore $2^7$ such solutions corresponding to the
different choices of the signs in $W\sub{2}$. However, since all seven scalars
are equivalent, only 8 solutions are distinct, namely the ones corresponding
to having $n$ $+$ signs and $7-n$ $-$ signs, with $n=0,\ldots,7$. However, the
solution where all signs are minus is covered by the second case, where
$W\sub{2}=\diag(w_-,\ldots, w_-)$ and $W\sub{3}$ is arbitrary. There are
therefore only 7 distinct solutions in the first class. For the second
class there is a unique choice for $W\sub{2}$, but $W\sub{3}$ is completely
arbitrary.\footnote{Note however that restrictions on $W\sub{3}$ can arise
as non-perturbative (in the scalar fields) effects. We will see how
this happens in an exactly solvable case below.}  Since $W\sub{3}$ is a
completely symmetric tensor of rank three,
it has $\frac{1}{3!}(N-1)N(N+1)=84$ independent components. There is
therefore an $84$-parameter family of solutions in this case.  Note that this
family is continuously connected to the supersymmetric solution
corresponding to (\ref{superpotential_0}) since $W_o$ also has
$W\sub{2}=\diag(w_-,\ldots, w_-)$. (See (\ref{Wo_Taylor}) below
for the Taylor expansion of $W_o$.)

All these solutions can be constructed systematically using the recursion
relations (\ref{recursion}). However, obtaining the solutions in closed form
by summing up the Taylor expansion is not very easy, if at all possible.
We can, however, obtain in closed form a subclass of these solutions by going
back to equation (\ref{potential}) and try to solve it exactly by first
reducing the number of dynamical scalar fields in a way that is consistent
with the equations of motion. A systematic way for doing this is setting some
of the eight scalar fields $X_i$ equal to each other in all possible ways.
Note that this is consistent with the equations of motion (\ref{eqs_motion}).
The independent ways to set a number of the scalars $X_i$ equal is to consider
all possible $n$-partitions of 8. Each $n$-partition corresponds to
an independent way to keep $n-1$ dynamical scalar fields.
Table \ref{partitions} lists all such partitions, together with the
corresponding isometry group \cite{Bakas:1999fa}.
\TABLE{
\begin{tabular}{|c||c|c|c|}
\hline
$n$ & partition of 8 & scalar fields & isometry group  \\
\hline
1 & 8 & 0 & $SO(8)$  \\
\hline
2 & 1+7 & 1 & $SO(7)$ \\
& 2+6 & & $SO(2)\times SO(6)$ \\
& 3+5 & & $SO(3)\times SO(5)$ \\
& 4+4 & & $SO(4)\times SO(4)$  \\
\hline
3 & 1+1+6 & 2 & $SO(6)$  \\
& 1+2+5 & & $SO(2)\times SO(5)$  \\
& 1+3+4 & & $SO(3)\times SO(4)$  \\
& 2+2+4 & & $SO(2)\times SO(2)\times SO(4)$  \\
& 2+3+3 & & $SO(2)\times SO(3)\times SO(3)$  \\
\hline
4 & 1+1+1+5 & 3 & $SO(5)$  \\
& 1+1+2+4 & & $SO(2)\times SO(4)$  \\
& 1+1+3+3 & & $SO(3)\times SO(3)$  \\
& 1+2+2+3 & & $SO(2)\times SO(2)\times SO(3)$  \\
& 2+2+2+2 & & $SO(2)\times SO(2)\times SO(2)\times SO(2)$ \\
\hline
5 & 1+1+1+1+4 & 4 & $SO(4)$  \\
& 1+1+1+2+3 & & $SO(2)\times SO(3)$  \\
& 1+1+2+2+2 & & $SO(2)\times SO(2)\times SO(2)$ \\
\hline
6 & 1+1+1+1+1+3 & 5 & $SO(3)$ \\
& 1+1+1+1+2+2 & & $SO(2)\times SO(2)$ \\
\hline
7 & 1+1+1+1+1+1+1+2 & 6 & $SO(2)$  \\
\hline
8 & 1+1+1+1+1+1+1+1+1 & 7 & - \\
\hline
\end{tabular}
\caption{The possible ways to reduce the number of dynamical scalar fields
$X_i$, by setting a number of these equal to each other, correspond
to the different partitions of 8. The resulting isometry group is also
shown.}
\label{partitions}}
We will attempt to find a closed form for the above solutions only for
the cases with a single dynamical field, however. As we will see,
even this seemingly innocuous case, requires considerable effort.

\subsection{Domain walls with a single scalar}
\label{one-scalar}

The four distinct one-scalar truncations in Table \ref{partitions} are obtained
by setting $X_1=\ldots=X_k\equiv X$, $X_{k+1}=\ldots=X_8=X^{-k/(8-k)}$, where
$k=4,5,6,7$. In this section we will keep $k$ as a parameter, however, so
that we can discuss all four cases simultaneously. The scalar potential
(\ref{potential2}), which for general $k$ takes the form\footnote{From now
on we drop the subscript $D$ in the gravitational constant $\k$ and the
AdS radius $l$ since we will always work in $D=4$.}
\be\label{1-scalar-pot}
V=-\frac{1}{16\k^2 l^2}\left(k(k-2)X^2+2k(8-k)X^{-2(k-4)/(8-k)}+
(8-k)(6-k)X^{-2k/(8-k)}\right),
\ee
is shown explicitly for each of the four cases in Table \ref{potentials}.
\TABLE{
\begin{tabular}{|c||c|c|c|}
\hline
k & isometry group & $-16\k^2l^2V$ & fixed points \\
\hline
7 & $SO(7)$ & $35 X^2+14 X^{-6}-X^{-14}$ &
$X=1,\;1/5^{1/8}$ \\&&&\\
6 & $SO(2)\times SO(6)$ & $24\left(X^2+X^{-2}\right)$ &
$X=1$ (double)\\&&&\\
5 & $SO(3)\times SO(5)$ & $3\left(5 X^2+X^{-10/3}
+10 X^{-2/3}\right)$ & $X=1$ \\&&&\\
4 & $SO(4)\times SO(4)$ & $8\left(X^2+X^{-2}+4\right)$ &
$X=1$ (double)\\&&&\\
\hline
\end{tabular}
\caption{The scalar potential for the four possible one-scalar
truncations. Note that the fixed point $X=1$, common to all potentials,
corresponds to the AdS fixed point at $\f=0$.}
\label{potentials}}
It is useful to parameterize the single scalar field $X$ in terms of a scalar
with a canonical kinetic term as
\be\label{X}
X=e^{\sqrt{\frac{8-k}{2k}}\k\f}.
\ee
Equation then (\ref{potential}) takes the form
\be\label{1-scalar-eq}
V=\frac{\k^2}{4}\left(\frac{(8-k)}{k}(X\pa_X W)^2-3W^2\right).
\ee
Moreover, the superpotential (\ref{superpotential_0}) becomes
\be\label{1-scalar-sup}
W_o=-\frac{1}{4\k^2 l}\left(k X+(8-k)X^{-\frac{k}{(8-k)}}\right),
\ee
and it is easily seen to be a solution of (\ref{1-scalar-eq}).

We have seen above that there exists a one-parameter family of functions
$W(\f;\a)$ which contains $W_o(\f)$ as a special case. In particular, the
Taylor expansions of $W(\f;\a)$ around $\f=0$, for a generic value of the free
parameter $\a$, and of $W_o(\f)$ have the same quadratic term, corresponding
to $w_-$ in the notation of the previous section.\footnote{Recall that the
parameter $\a$ first enters in the cubic term in the Taylor expansion around 
$\f=0$.}
In addition, however, there exists another isolated solution,
$\widetilde{W}_o(\f)$, whose quadratic term corresponds to $w_+$ and
whose cubic term vanishes. The Taylor expansions of $W(\f;\a)$ and
$\widetilde{W}_o(\f)$ around $\f=0$ are therefore not continuously connected.
This though does not exclude the possibility that, {\em non-perturbatively} in
$\f$, $W(\f;\a)$ and $\widetilde{W}_o(\f)$ are continuously connected.
Remarkably, we will see below in an example where the exact one-parameter
family $W(\f;\a)$ can be obtained exactly that $W(\f;\a)$ interpolates
between the supersymmetric solution $W_o(\f)$ and $\widetilde{W}_o(\f)$.

In the next section we will address systematically the problem of solving
equation (\ref{1-scalar-eq}) exactly. For the moment, however, we can
use the fact that $W(\f;\a)$ is continuously connected to $W_o(\f)$ in order
to obtain $W(\f;\a)$ in an expansion in the free parameter $\a$, for
general $k$. Obviously, this approach can provide no information on
$\widetilde{W}_o(\f)$. We start by writing $W(\f;\a)$ in a formal
asymptotic expansion as\footnote{The normalization of the free parameter is
chosen so that it matches the natural free parameter of the exact solution
that we will present in the next section for $k=6$.}
\be
W(\f;\a)=W_o(\f)+\sum_{n=1}^\infty \left(\frac{-1}{32\k^2l}\right)^n
(\a-\a_o)^nW^{(n)}(\f),
\ee
where, $\a_o=-(8-k)(k-4)k/24$ and  $W(\f;\a_o)\equiv W_o(\f)$. Inserting this
expansion into (\ref{1-scalar-eq}) one obtains an infinite set of
{\em linear} equations for the functions $W^{(n)}(\f)$, namely
\bea\label{1-scalar-recurs}
&&\left(X\pa_X W_oX\pa_X -\frac{3k}{(8-k)}W_o\right)W^{(n)}\NO\\
&&+\frac12\sum_{m=1}^{n-1}\left(X\pa_X W^{(m)}X\pa_X W^{(n-m)}
-\frac{3k}{(8-k)}W^{(m)}W^{(n-m)}\right)=0,
\eea
which can be solved iteratively. For $n=1$, this equation is homogeneous
and its solution is
\be
W^{(1)}=\left(\frac{X^{\frac{8}{(8-k)}}-1}{X}\right)^3.
\ee
Note that, as expected from the general analysis above, $W^{(1)}=\co(\f^3)$
as $\f\to 0$. For $n>1$ equation (\ref{1-scalar-recurs}) is non-homogeneous
but it can be solved with the help of an integrating factor
\be
R=\exp\left(-\frac{3k}{(8-k)}\int \frac{dX}{X^2}\frac{W_o}{\pa_X W_o}\right)=
\frac{1}{W^{(1)}}.
\ee
The solution then takes the form
\be
W^{(n)}=W^{(1)}\int\frac{dX}{X}\frac{Q_n(X)}{W^{(1)}(X)}+c_n W^{(1)},
\ee
where $c_n$ are constants and
\be
Q_n=-\frac{1}{2X\pa_X W_o}\sum_{m=1}^{n-1}\left(X\pa_X W^{(m)}X\pa_X W^{(n-m)}
-\frac{3k}{(8-k)}W^{(m)}W^{(n-m)}\right).
\ee
In particular,
\be
W^{(2)}=6\k^2l \left(\frac{X^{\frac{8}{(8-k)}}-1}{X}\right)^3
\left(\frac{1}{(8-k)}X^{\frac{4(k-2)}{(8-k)}}+\frac{2}{(k-4)}
(X^{\frac{4(k-4)}{(8-k)}}-1)+\frac1kX^{\frac{4(k-6)}{(8-k)}}+c_2(k)\right),
\ee
where the term involving $k-4$ in the denominator is understood as the limit
$k\to 4$, giving $\log X$,  for the case $k=4$. Moreover, the constant
$c_2(k)$ is not arbitrary. It is uniquely fixed by the requirement that
$W^{(2)}$ does not contribute to the cubic term in $\f$ of $W(\f;\a)$,
which is necessary in order to identify $(\a-\a_o)$ (as opposed to some other
function of $\a$) with the free parameter of $W$.
\be
c_2(k)=-\frac{8}{k(8-k)}.
\ee
The same argument determines all constants $c_n$ in $Q_n$. Putting
everything together, to this order we have
\bea\label{pert-superpotential}
W(\f;\a) & = & -\frac{1}{4\k^2 l}\left(k X+(8-k)X^{-\frac{k}{(8-k)}}\right)
\\&&-\frac{1}{32\k^2l}(\a-\a_o)\left(\frac{X^{\frac{8}{(8-k)}}
-1}{X}\right)^3\left\{1-\frac{3}{16}(\a-\a_o)
\left(\frac{1}{(8-k)}X^{\frac{4(k-2)}{(8-k)}}\right.\right.\NO\\
&&\left.\left.+\frac{2}{(k-4)}(X^{\frac{4(k-4)}{(8-k)}}-1)
+\frac1kX^{\frac{4(k-6)}{(8-k)}}-\frac{8}{k(8-k)}\right)
\right\}+\co\left((\a-\a_o)^3\right).\NO
\eea

Given this perturbative (in $\a-\a_o$) fake superpotential, we can immediately
obtain the corresponding domain wall solutions via the first order equations
(\ref{flow_eqs}). We give explicitly the form of these backgrounds to
first order in $\a-\a_o$ in Appendix \ref{X-coord}, since we will need them
for the computation of the one- and two-point functions of the field
theory duals of these domain walls.

\subsection{Exact closed form solutions}
\label{k=6}

Having obtained a perturbative solution for $W(\f;\a)$ for all possible values
of $k$, let us now try to solve (\ref{1-scalar-eq}) exactly. This should
determine not only the full  $W(\f;\a)$, but also $\widetilde{W}_o(\f)$.
It was observed in \cite{Papadimitriou:2004ap} that for a single
scalar field, $\f$, equation (\ref{potential}), with an arbitrary potential,
can be recast in a standard form by means of the field redefinitions
\be\label{field-redef}
\psi=\sqrt{\frac{d\k^2}{d-1}}\f,\quad y=\coth(u),\quad W=l v\cosh(u),
\ee
where
\be\label{pot_param}
v=-\left(-\frac{2(d-1)}{d\k^2l^2}V\right)^{1/2}.
\ee
In terms of these variables, equation (\ref{potential})  takes the
form\footnote{Note that the obvious solutions $y=\pm 1$ of this equation
are rejected since, via (\ref{field-redef}), they correspond to $u\to\infty$
and hence $W\to\infty$.}
\be\label{Abel-special}
y'(\psi)=\left(\frac{v'}{v}y-1\right)(y^2-1),
\ee
where the prime denotes derivative with respect to $\psi$. This equation
is a special case of Abel's equation of the first kind \cite{Kamke}
\be\label{Abel}
y'=f_3(\psi)y^3+f_2(\psi)y^2+f_1(\psi)y+f_0(\psi),
\ee
where $f_i(\psi)$ are arbitrary functions. Abel's equation can in turn be cast
in the canonical form
\be\label{Abel-canonical}
z'=\tilde{f}_3(\psi)z^3+\tilde{f}_1(\psi)z
+\tilde{f}_0(\psi),
\ee
by means of the transformation
\be
y=z-\frac{f_2}{3f_3}.
\ee
Clearly, equation (\ref{Abel-canonical}) can be integrated directly if either
$\tilde{f}_3$ or $\tilde{f_0}$ vanish. Moreover, it can also be integrated
directly if `Abel's invariant'
\be
\ci\equiv -\frac{\left(\tilde{f}_0\tilde{f}'_3-\tilde{f}'_0\tilde{f}_3
+3\tilde{f}_0\tilde{f}_3\tilde{f}_1\right)^3}{27\tilde{f}_3^4\tilde{f}_0^5},
\ee
is a constant \cite{Kamke}. If it is not a constant, however, no general
solution of (\ref{Abel-canonical}) is known. In that case one can only hope
that the equation at hand falls into one of the known integrable classes of
Abel's equation, each of which has a very particular way of solution that is
not applicable to other classes. Some recent investigations and overviews
of Abel's equation and its known integrable classes can be found in
\cite{Abel,Kamke}.

In our case, however, the functions $f_i(\psi)$ are not completely arbitrary
since they are all related to the scalar potential. Specifically, from
(\ref{Abel-special}) we read
\be
f_3=-f_1=q',\qquad f_2=-f_0=-1,
\ee
where $q\equiv \log|v|$. Moreover, one can easily compute the tilded
coefficients corresponding to the transformed equation (\ref{Abel-canonical}):
\be
\tilde{f}_3=q',\quad \tilde{f}_1=-\frac{1}{3q'}
(1+3q'^2),\quad \tilde{f}_0=\frac{1}{3q'^2}(q''+2q'^2-2/9).
\ee
It follows that for a generic potential, and hence a generic $q$, Abel's
invariant is not automatically constant. Requiring that it be a constant,
leads to a second order, non-linear differential equation for $q'$, which
seems more difficult to solve than the original first order equation.
However, as we have already pointed out, requiring that either
$\tilde{f}_3$ or $\tilde{f}_0$ vanish, also leads to a solvable equation.
These conditions lead to differential equations for the potential, which are
easily solvable. In particular, $\tilde{f}_3=q'=0$
gives the constant potential
\be
V=-\frac{d(d-1)}{2\k^2 l^2},
\ee
corresponding to exact AdS space. More interesting is the condition
$\tilde{f}_0=\frac{1}{3q'^2}(q''+2q'^2-2/9)=0$, which leads to the potential
\be\label{potential_1}
V=-\frac{d(d-1)}{2\k^2l^2}\cosh\left(\frac{2\psi}{3}\right).
\ee
The observation that this potential leads to a soluble Abel's equation
was the only motivation for considering this potential in
\cite{Papadimitriou:2004ap}. Curiously, however, noting from (\ref{X}) and
(\ref{field-redef}) that, for $d=3$, $X$ and $\psi$ are related by
\be\label{X-psi}
X=e^{\sqrt{\frac{8-k}{3k}}\psi},
\ee
the potential (\ref{potential_1}) is seen to be identical to
the potential (\ref{1-scalar-pot}) for $k=6$. Hence, at least for
the case $k=6$, we are able to solve (\ref{1-scalar-eq}) exactly.
However, the potential (\ref{1-scalar-pot}) with generic $k$ gives
\be
q'=\frac{v'}{v}=\sqrt{\frac{k(8-k)}{3}}\left(\frac{(k-2)e^{16\psi/
\sqrt{3k(8-k)}}-2(k-4)e^{8\psi/\sqrt{3k(8-k)}}-(6-k)}
{(k-2)e^{16\psi/\sqrt{3k(8-k)}}+2(8-k)e^{8\psi/\sqrt{3k(8-k)}}+(6-k)(8-k)}
\right).
\ee
One can now easily check that, except for $k=6$ in which case
$\tilde{f}_0$ vanishes, Abel's invariant is not constant for any value
of $k$. As we discussed already, this makes it much harder to solve
(\ref{Abel-special}) for the potential (\ref{1-scalar-pot}) with
$k\neq6$.

To obtain the exact solution for the case $k=6$, we start by inserting the
potential (\ref{potential_1}) in equation (\ref{Abel-special}). The resulting
equation takes the form
\be
\frac{2}{1-s^2}\frac{ds}{d\r}+\frac{1}{1-\r^2}\left(\frac{\r}{s}-3\right)=0,
\ee
where
\be\label{coord-def}
s=\frac{1}{y},\qquad \r=\tanh\left(\frac{2\psi}{3}\right).
\ee
The general solution of this equation is \cite{Papadimitriou:2004ap}
\be\label{exact-sol}
s=\frac{\r}{1\pm(1-\r^2)(1+2\a\r+\r^2)^{-1/2}},
\ee
where $\a$ is an integration constant. Since the conformal boundary corresponds
to $\r=0$, we can take $\r\geq 0$. The choice $\r\leq 0$ is also possible but
it is equivalent. The value of the integration constant $\a$
is then restricted by the requirement that $1+2\a\r+\r^2\geq 0$. This is
guaranteed provided
\be
\a\geq -1.
\ee

The fake superpotential is now obtained from (\ref{field-redef}) as
\bea\label{exact-fake-super}
W(\f;\a)=-\frac{2}{\k^2l}\frac{1}{(1-\r^2)^{1/4}}\frac{1}{\sqrt{1-s^2}}.
\eea
Expanding this for small $\psi$, we see that the solution with the
negative sign in (\ref{exact-sol}) always contains a linear term in $\psi$
and it is therefore rejected. For the positive sign solution we find
\be
W(\f;\a)=-\frac{2}{\k^2l}\left(1+\frac 16\psi^2+\frac{1}{27}\a\psi^3+
\co(\psi^4)\right),
\ee
which is precisely of the required form. We therefore expect that this
is the full one-parameter family of fake superpotentials whose existence
we predicted above on general grounds and which we computed perturbatively
in the free parameter. In particular, it should contain the true
superpotential (\ref{1-scalar-sup}), which for $k=6$ becomes
\be
W_o(\f)=-\frac{1}{2\k^2l}\left(3e^{\psi/3}+e^{-\psi}\right).
\ee
Indeed, this is the case as it is easy to check that for $\a=-1$, $W(\f;\a)$
reduces to $W_o(\f)$:
\be
W(\f;-1)=W_o(\f).
\ee
Since $W(\f;\a)$ is the most general solution, however, one wonders where
is the solution $\widetilde{W}_o(\f)$ which we have predicted and whose
expansion around $\f=0$ should have a different quadratic term from that
of $W(\f;\a)$. The answer is that $\widetilde{W}_o(\f)$ is obtained from
$W(\f;\a)$ by sending $\a$ to infinity:
\be
\widetilde{W}_o(\f)=\lim_{\a\to\infty}W(\f;\a)=-\frac{2}{\k^2l}\cosh^{3/2}
\left(\frac{2\psi}{3}\right).
\ee
Expanding this for small $\psi$ we find
\be
\widetilde{W}_o(\f)=-\frac{2}{\k^2l}\left(1+\frac13\psi^2+\co(\psi^4)\right).
\ee
This has precisely the desired form, namely a quadratic term corresponding
to $w_+$ and a vanishing cubic term. The fake superpotential $W(\f;\a)$,
therefore, interpolates between the supersymmetric superpotential
$W_o(\f)=W(\f;-1)$ and $\widetilde{W}_o(\f)=W(\f;\infty)$.

\section{Exact non-supersymmetric membrane flows}
\label{uplift}

All non-supersymmetric domain wall solutions we have obtained
above in $D=4$, in closed form or not, can in principle be uplifted to
asymptotically $AdS_4\times S^7$ non-supersymmetric solutions of
eleven-dimensional supergravity using the ansatz (\ref{ansatz}).
We will only uplift explicitly the closed form solutions we
found in the previous section, however. To do this we first need to
determine the four-dimensional domain wall metrics corresponding
to the exact fake superpotentials for $k=6$.

Integrating the first order equations (\ref{flow_eqs}) using
the fake superpotential (\ref{exact-fake-super}) we find that the
full one-parameter family of Poincar\'e domain walls takes the form
\bea\label{metric-a}
ds^2_\a&=&\frac{1}{2\r^2}\left(1+\a\r+\sqrt{1+2\a\r+\r^2}\right)
\left(\frac{l^2d\r^2}{\sqrt{1-\r^2}(1+2\a\r+\r^2)}+\x^2\sqrt{1-\r^2}\;
\h_{ij}dx^idx^j\right),\NO\\\NO\\
\f&=&\sqrt{\frac{3}{2\k^2}}\tanh^{-1}\r.
\eea
The integration constant $\x^2$ can be absorbed by a rescaling of the
transverse coordinates $x^i$, but we have introduced it for reasons that
will become clear soon. Namely, for any finite value of $\a$, and taking
$\x^2=1$, the metric (\ref{metric-a}) is asymptotically AdS with canonical
radial coordinate $\r\sim e^{-r/l}$ as $\r\to 0$. In particular, the
supersymmetric metric corresponding to $\a=-1$ reads
\be\label{metric-1}
ds^2_{-1}=\frac{1}{\r^2}\left(\frac{l^2d\r^2}{\sqrt{1+\r}(1-\r)^{3/2}}+
\sqrt{1+\r}(1-\r)^{3/2}\h_{ij}dx^idx^j\right).
\ee
In order for the metric (\ref{metric-a}) to have a well-defined limit as
$\a\to\infty$, however, we must take $\x^2\sim {\rm const.}/\a$ as
$\a\to\infty$. Taking $\x^2\sim 2/\a$ and evaluating the limit $\a\to\infty$,
(\ref{metric-a}) becomes
\be\label{metric-infty}
ds^2_\infty=\frac{l^2d\r^2}{4\r^2\sqrt{1-\r^2}}+
\frac{\sqrt{1-\r^2}}{\r}\;\h_{ij}dx^idx^j.
\ee
This is again an asymptotically AdS metric, but with canonical radial
coordinate $\sqrt{\r}\sim e^{-r/l}$.

Note that these metrics are non-singular for $0\leq\r<1$. There is a
singularity at $\r=1$, however, which is in fact a curvature
singularity from the four-dimensional point of view and the scalar field also
diverges at this point. In fact, the curvature singularity of the
supersymmetric metric, for which the Ricci scalar behaves like
$R_4\sim (1-\r)^{-1/2}$ as $\r\to1$, is milder that the curvature singularity
of the non-supersymmetric metrics, for which $R_4\sim (1-\r)^{-3/2}$ as
$\r\to1$. Moreover, the singularity is null for the supersymmetric
case but timelike for the non-supersymmetric metric \cite{Gubser:2000nd}.
Nevertheless, in both cases the singularity is `good' according
to the criterion of \cite{Gubser:2000nd} since the scalar potential
(\ref{potential_1}) is bounded from above, not only on-shell but even
off-shell. Accordingly, in both cases, the presence of the singularity
signals some genuine IR phenomenon in the dual field theory.

We can now use the ansatz (\ref{ansatz}) to uplift the four-dimensional
solution (\ref{metric-a}) to eleven dimensions. It is convenient, however,
to first use a reduced ansatz obtained from (\ref{ansatz}) by setting the
8 scalars $X_i$ pairwise equal \cite{Cvetic:1999xx}:
\be
X_{2a-1}=X_{2a}\equiv\tilde{X}_a,\quad a=1,2,3,4,
\ee
so that $\tilde{X}_1\tilde{X}_2\tilde{X}_3\tilde{X}_4=1$.
This reduction corresponds to the scalar sector of the truncation
of $\cn=8$ supergravity to the maximal abelian subgroup, $U(1)^4$, of its
gauge group $SO(8)$ \cite{Duff:1999gh,Cvetic:1999xp}.\footnote{The $U(1)$
gauge fields and the three axions are set to zero here, however.}
Note, that this reduction does not include all possible one-scalar truncations
discussed in Section \ref{one-scalar} since the cases $k=3,5$ are not
consistent with this reduction. It does cover however the cases $k=4$ and
$k=6$, which is the case we are interested here. The reduced ansatz reads
\bea\label{ansatz-reduced}
d\hat{s}_{11}^2&=&\tilde{\D}^{2/3}ds_4^2+4l^2\tilde{\D}^{-1/3}\sum_{a=1}^4
\tilde{X}_a^{-1}\left(d\tilde{\m}_a^2+\tilde{\m}_a^2d\f_a^2\right),\\
\hat{F}\sub{4}&=&\frac{1}{l}\sum_{a=1}^4\left(\tilde{X}_a^2\tilde{\m}_a^2-
\tilde{\D}\tilde{X}_a\right)\e\sub{4}
-l\sum_{a=1}^4\tilde{X}_a^{-1}\ast d\tilde{X}_a\wedge d(\tilde{\m}_a^2),
\eea
where $\tilde{\D}=\sum_{a=1}^4\tilde{X}_a\tilde{\m}_a^2$ and the quantities
$\tilde{\m}_a$ and the four angles $\f_a$, $0\leq \f_a\leq 2\p$,
are related to the direction cosines $\m_i$ in (\ref{ansatz}) by
\be
\m_{2a-1}=\tilde{\m}_a\cos\f_a,\qquad \m_{2a}=\tilde{\m}_a\sin\f_a,\qquad
a=1,2,3,4,
\ee
so that $\sum_{a=1}^4\tilde{\m}_a^2=\sum_{i=1}^8\m_i^2=1$. The four
$\tilde{\m}_a$ can be parameterized in terms of the angles on a three-sphere as
\be
\tilde{\m}_1=\cos\th\cos\chi\cos\om,\quad \tilde{\m}_2=\cos\th\cos\chi\sin\om,
\quad \tilde{\m}_3=\cos\th\sin\chi,\quad \tilde{\m}_4=\sin\th,
\ee
$0\leq \th,\chi\leq \p$, $0\leq \om\leq 2\p$. Finally, the four scalars
$\tilde{X}_a$ can be parameterized in terms of three dilatonic scalars
$\vec{\tilde{\vf}}=(\tilde{\vf}_1,\tilde{\vf}_2,\tilde{\vf}_3)$:
\be
\tilde{X}_a=e^{-\frac12\vec{\tilde{b}}\cdot\vec{\tilde{\vf}}},
\ee
where $\vec{\tilde{b}}_i$ satisfy
\be
\vec{\tilde{b}}_a\cdot\vec{\tilde{b}}_b=4\d_{ab}-1.
\ee
A convenient choice for $\vec{\tilde{b}}_a$ is
\be
\vec{\tilde{b}}_1=(1,-1,-1),\quad \vec{\tilde{b}}_2=(-1,1,-1),\quad
\vec{\tilde{b}}_3=(-1,-1,1),\quad \vec{\tilde{b}}_4=(1,1,1).
\ee

The $k=6$ solution now corresponds to setting
$\tilde{X}_1=\tilde{X}_2=\tilde{X}_3\equiv X$, $\tilde{X}_4=X^{-3}$.
Recalling from (\ref{X-psi}) that $X=e^{\psi/3}$ and the relation
between $\psi$ and $\r$ from (\ref{coord-def}), we deduce that
\be
X=\left(\frac{1+\r}{1-\r}\right)^{1/4}.
\ee
Hence,
\be
\tilde{\D}=\left(\frac{1+\r}{1-\r}\right)^{1/4}\cos^2\th+
\left(\frac{1-\r}{1+\r}\right)^{3/4}\sin^2\th.
\ee

Putting everything together, we have the following two solutions of
eleven-dimensional supergravity
\bea\label{11-metric}
d\hat{s}_{11}^2&=&\tilde{\D}^{2/3}ds_4^2+4l^2\tilde{\D}^{-1/3}
\left\{\left(\frac{1+\r}{1-\r}\right)^{3/4}\left[\left(\cos^2\th+
\left(\frac{1-\r}{1+\r}\right)\sin^2\th\right)d\th^2+\sin^2\th d\f_4^2\right]
\right.\NO\\
&&\left.+\left(\frac{1-\r}{1+\r}\right)^{1/4}\cos^2\th d\Om_5^2\right\},
\eea
where $ds_4^2$ is given either by the finite-$\a$ metric (\ref{metric-a}) or
by the $\a\to\infty$ metric (\ref{metric-infty}). Correspondingly, the
four-form field strength is given by
\bea\label{F-a}
\hat{F}^\a\sub{4}&=&\frac{(1+\a\r+\sqrt{1+2\a\r+\r^2})}
{2\r^2{\sqrt{1+2\a\r+\r^2}}}\NO\\
&&\left\{\frac{(1+\r)}{2\r^2}\left[2\cos^2\th+\left(\frac{1-\r}{1+\r}\right)
(1+2\sin^2\th)\right](1+\a\r+\sqrt{1+2\a\r+\r^2})
d\r\right.\NO\\&&\left.
+4(1+2\a\r+\r^2)\cos\th\sin\th d\th\rule[.2cm]{0cm}{.3cm}\right\}
\wedge\bar{\e}\sub{3},\\\NO\\
\label{F-infty}
\hat{F}^\infty\sub{4}&=&\frac{1}{\sqrt{\r}}\left\{\frac{(1+\r)}{2\r^2}
\left[2\cos^2\th+\left(\frac{1-\r}{1+\r}\right)(1+2\sin^2\th)\right]d\r
+8\cos\th\sin\th d\th\right\}\wedge\bar{\e}\sub{3}.\NO\\
\eea
It is not difficult to check that these satisfy ${\rm d}\hat{F}\sub{4}=0$
and ${\rm d}\hat{\ast}\hat{F}\sub{4}=0$. Of course, (\ref{11-metric})
and (\ref{F-a})-(\ref{F-infty}) also satisfy Einstein's equation in eleven
dimensions, as is guaranteed by the fact that the four-dimensional theory
is a consistent truncation of eleven-dimensional supergravity
\cite{Cvetic:2000eb}. We have not checked this explicitly, however.

A few comments are in order here. First, note that the compact part the
metric (\ref{11-metric}) does not depend on the parameter $\a$ and hence
it describes the same inhomogeneous deformation of $S^7$ as the
supersymmetric solution with $\a=-1$. Namely, at $\r=0$ the compact part of
the metric is exactly the metric on $S^7$. As one moves away from $\r=0$
the $S^7$ is deformed to a warped product of an $S^5$ of decreasing radius
and a squashed $S^2$ with increasing radius. At $\r=1$, the $S^5$ shrinks to
zero size, while the $S^2$ becomes totally squashed, but with infinite radius.
The supersymmetric solution corresponds to a continuous non-uniform
distribution of $M2$-branes on a disc of finite radius on the equatorial
plane of the squashed $S^2$ \cite{Kraus:1998hv, Cvetic:1999xx}. It would be
very interesting to find an analogous interpretation for the
non-supersymmetric solutions. As the four-dimensional solutions, the uplifted
metrics have a curvature singularity at $\r=1$, but now the eleven-dimensional
Ricci scalar behaves like $\hat{R}=\frac16 \hat{F}\sub{4}^2\sim (1-\r)^{-1/3}$
as $\r\to1$, {\em independently} of the value of $\a$. Of course, the 
singularity remains null for the supersymmetric case and timelike for the
non-supersymmetric one since the uplift does not alter the causal structure.
However, at least for the supersymmetric solution, the uplift helps
identify the cause of the singularity, namely the fact that the distribution
of the $M2$-branes is continuous \cite{Kraus:1998hv}, and as a result
understand how M-theory resolves the singularity. A similar
interpretation for the non-supersymmetric solution would therefore
clarify the nature of the singularity. Another important
difference between the supersymmetric and non-supersymmetric solutions is
that the part of the metric orthogonal to the squashed $S^2$ becomes
conformal to $AdS_4\times S^5$ as $\r\to0$ for the supersymmetric case, while
for the non-supersymmetric case it becomes
conformal to $\mathbb{R}^4\times S^5$ as $\r\to 0$. Moreover,
$\hat{F}\sub{4}$ vanishes at $\r= 0$ for the supersymmetric case, while it
is finite but non-zero for the non-supersymmetric one.

Interestingly, the same scalar potential (\ref{potential_1}),
corresponding to $k=6$, led to the MTZ black hole
in four dimensions \cite{Martinez:2004nb}. Since we know how to uplift
solutions of the four-dimensional scalar-gravity system with this potential
to eleven dimensions, we find it tempting to present the
eleven-dimensional black hole metric explicitly, which we do in Appendix
\ref{MTZ}.

\section{Holographic one-point functions}
\label{1-pt-fns}

The asymptotically AdS domain walls (\ref{poincare}) describe, via the
AdS/CFT duality, the RG flow of the field theory living on the conformal
boundary. Such an RG flow can result from a deformation of the Lagrangian
of the UV CFT by a relevant operator, or from a non-conformal vacuum,
described by the VEVs of certain operators. To determine which of these
possibilities is realized in a given domain wall background, one should
evaluate holographically the one-point functions of the operators
dual to the non-trivial scalar fields, as well as the one-point function
of the stress tensor.

One, therefore, first needs to identify the gauge-invariant operator
$\co_\D$ dual to a given scalar field. Recall that the mass, $m$, of a scalar
field is related to the dimension, $\D$, of the dual operator via
\be
m^2l^2=\D(\D-d).
\ee
Since this equation has two roots, $\D_\pm$, the question arises
as to which of the two is the dimension of the dual operator. It was
argued in \cite{Klebanov:1999tb} that while for $m^2l^2>-(d/2)^2+1$ the
dual operator must unambiguously have dimension $\D_+$, for
\be\label{two-quant-condition}
 -\left(\frac d2\right)^2\leq m^2l^2\leq -\left(\frac d2\right)^2+1,
\ee
both $\D_\pm$ are possible dimensions for the dual operator. More
specifically, there are two possible quantizations of the scalar field,
corresponding to the two dimensions $\D_\pm$ of the dual operator
\cite{Breitenlohner:1982jf}. The resulting generating functionals of
correlation functions of the corresponding operators are then related by a
Legendre transformation as we will review below.

An important property of the $SL(N,\mathbb{R})/SO(N)$ scalars is
that their mass falls precisely in the range (\ref{two-quant-condition})
allowing two quantizations. Namely, recall from (\ref{V_Taylor})
that the mass of the scalar fields of the $SL(N,\mathbb{R})/SO(N)$ sector is
\be
m_I^2l_D^2=2(2-d).
\ee
With this mass, the condition (\ref{two-quant-condition}) translates
into\footnote{Curiously, this is precisely the range of dimensions for
which there exist superconformal quantum field theories.}
\be
 2\leq d\leq 6,
\ee
which includes all cases for $d$ we are interested in, namely $d=3,4,6$.
The two possible dimensions are
\be\label{deltas}
\D_\pm=\frac d2\pm\frac12|d-4|,
\ee
which coincide for $d=4$. In this case the mass saturates the BF bound
$m^2l^2\geq -(d/2)^2$, and there is a unique quantization
\cite{Klebanov:1999tb}. For $d=3$ or $d=6$, however, there are two
possible quantizations and consequently two possibilities
for the dimension of the dual operators. As we have discussed, however, for
$d=6$ this ambiguity is removed by symmetry, which determines that the 
dual operators have dimension $\D_+=4$. But we are interested in the
case $d=3$ here, which is the only case admitting non-supersymmetric
fake superpotentials, and since there is an ambiguity in this case we will
analyze the two possible quantizations separately. We will keep the analysis 
and the notation as general as possible, though, so that the analysis is 
applicable to other cases too.

Let us start by recalling that the asymptotic form of the potential
(\ref{V_Taylor}) implies that a generic solution to the bulk scalar field
equation of motion takes the form
\be\label{generic-scalar}
\f(r,x)\sim e^{-\D_-r/l}(\f_-(x)+\cdots)+e^{-\D_+r/l}(\f_+(x)+\cdots).
\ee
Since we are excluding the case where the BF bound is saturated, we
have $\D_-<\D_+$ and so the term involving $\f_-$ dominates
asymptotically as $r\to\infty$. For a particular solution, however,
such as a domain wall of the form (\ref{poincare}), one of the
functions $\f_\pm$ can be zero. This depends entirely on the fake
superpotential $W(\f)$ that defines the flow equations (\ref{flow_eqs}).
To determine the VEV of the operator dual to the scalar field $\f$, one
should evaluate the bulk action on the solution (\ref{generic-scalar}),
which is identified with the generating functional of correlation
functions of the dual operator \cite{Klebanov:1999tb}.

Now, the on-shell action evaluated on a (Euclidean) Poincar\'e domain wall
(\ref{poincare}) is \cite{dBVV,Bianchi:2001de,Papadimitriou:2004rz}
\be
S_{\rm on-shell}^B=\int d^dx \sqrt{\g_B}W(\f_B),
\ee
where $\g_{Bij}=e^{2A}\d_{ij}$ and $W(\f_B)$ is the fake superpotential
that defines the flow equations (\ref{flow_eqs}). We have included the
subscript $B$ here to emphasize that this is the on-shell action evaluated
on the background domain wall solution. We will need to consider fluctuations
around this background when we calculate two-point functions later on.
As is well known, however, the on-shell action diverges and one needs to
remove this divergence by adding covariant counterterms
\cite{Henningson:1998gx, Balasubramanian:1999re, Kraus:1999di,dBVV,
deHaro:2000xn, Bianchi:2001kw, Martelli:2002sp, Papadimitriou:2004ap}. Although
the covariant counterterms are a property of the supergravity action
(\ref{fake_action}) -  that is, once constructed in full generality
by the asymptotic analysis of the action (\ref{fake_action}), they
remove the divergences of the on-shell action when evaluated on {\em all}
extrema of (\ref{fake_action}) - for domain wall backgrounds of the form
(\ref{poincare}) they take particularly simple form, which can be
determined without the need to first compute the counterterms in full
generality. In particular, the part of the the counterterm action that
involves only the scalar fields, i.e. excluding the gravitational
counterterms (except from the volume renormalization which can be counted
with the scalar fields) and terms involving derivatives
of the scalars (which vanish on the domain wall background), are given by a
function $U(\f)$ that satisfies equation (\ref{potential}) at least
asymptotically \cite{dBVV,Martelli:2002sp,Papadimitriou:2004rz}, and has an
expansion
\be\label{U}
U(\f)=-\frac{d-1}{\k^2l}-\frac{1}{2l}\D_-\f^I\f^I+\co(\f^3).
\ee
The first term in this function is nothing but the well-known volume
renormalization term. The quadratic term requires some explanation, however.

Recall that, since $U(\f)$ satisfies (\ref{potential}) and the potential
has a Taylor expansion of the form (\ref{V_Taylor}),
$U(\f)$ has an expansion of the form (\ref{W_Taylor}) with the quadratic term
being a diagonal matrix with diagonal elements $w_\pm=-\D_\pm/2l$ (see Section
\ref{domain-walls}).\footnote{Note that for $d>4$, however,
$w_\pm=-\D_\mp/2l$. Here we are primarily interested in the case $d=3$.} There
are $2^n$ such matrices, where $n$ is the number of independent scalars. But
as we will now explain, there is a unique choice for the counterterms since
they must remove the divergences for {\em any} fake superpotential $W(\f)$,
whose quadratic term can indeed be any of these $2^n$ matrices. It suffices
to consider the case of a single on-shell scalar field, which takes the form
(\ref{generic-scalar}). If $\f_-\neq 0$, then its leading asymptotic behavior
is $\f\sim e^{-\D_- r/l}\f_-$, and so, by the flow equations (\ref{flow_eqs}),
the corresponding fake superpotential should have a quadratic term
with coefficient $-\D_-/2l$. Since this is the same quadratic term as
that of the counterterm $U(\f)$, the quadratic term in the on-shell action
will be canceled. In fact, one can take $U(\f)$ to be the fake superpotential
in this case - although this may not be necessary if there are no higher
order divergences. It is crucial though that this {\em same} counterterm
$U(\f)$ removes the divergences for the case when $\f_-$ vanishes, since
the counterterms are valid for {\em any} solution to a given bulk action.
In this case $\f\sim e^{-\D_+ r/l}\f_+$ asymptotically and so the
fake superpotential should have a quadratic term proportional to $-\D_+/2l$.
This means that upon subtracting the counterterm $U(\f)$, there will
be a quadratic term $-\frac{1}{2l}(\D_+-\D_-)\f^2$ left in the action.
However, $\f^2=\co(e^{2\D_+ r/l})$, in this case, and since $2\D_+>d$,
this term is not divergent and will drop out of the on-shell action as the
regulator is removed. This argument explains why $\D_-$ has to appear
in the quadratic term of the counterterm. Generalizing this argument
to more than one fields,\footnote{We assume that the scalar fields all have
the same mass squared, but the argument generalizes in an obvious way to
unequal masses.} the counterterm must have a quadratic term
proportional to the unit matrix with coefficient $-\D_-/2l$.

Let us now apply this to the case we are interested in.  From (\ref{deltas})
we have
\be
\D_-=d-2, \quad d<4, \qquad \D_-=2,\quad d>4.
\ee
Expanding the true superpotential (\ref{superpotential_0}) we get
\be\label{Wo_Taylor}
W_o(\f)=-\frac{(d-1)}{\k_D^2l_D}-\frac{(d-2)}{2l_D}\f^I\f^I+
\frac{(d-2)\sqrt{2}\k_D}{96l_D}\sum_{i=1}^Nb_{iI}b_{iJ}b_{iK}\f^I\f^J\f^K
+\co(\f^4).
\ee
It follows that, for $d<4$, we can use $W_o(\f)$ as the counterterm $U(\f)$:
\be
U(\f)=W_o(\f).
\ee
A few comments are in order here. At first sight, it seems that for $d=6$ we
are not able to use $W_o(\f)$ as the counterterm since it has the wrong
quadratic term, which is surprising since we know that $W_o(\f)$ corresponds
to a supersymmetric domain wall and, hence, one should be able to choose a
supersymmetric renormalization scheme where the on-shell action is identically
zero. The answer is that, as we showed in Section \ref{domain-walls}, for
$d=6$, the potential (\ref{V_Taylor}) requires that the quadratic term of
{\em any} fake superpotential is 
$\diag(-\D_+/2l,\ldots,-\D_+/2l)$.\footnote{This fact, in combination with the
fact that the dimension of the dual operators is unambiguously determined
to be $\D_+=4$, implies that all domain walls for the $SL(5,\mathbb{R})/SO(5)$
scalars in seven dimensions necessarily describe VEVs of the dual theory.}
Hence, there are simply no solutions with non-zero $\f_-$ in this case and so
$W_o(\f)$ can be safely used as the counterterm, resulting in the
expected supersymmetric renormalization scheme. Second, focusing on the
case $d=3$ which we are interested in, we have seen that there is a continuous
family of fake superpotentials, of the generic form
\be\label{W_Taylor_special}
W(\f)=-\frac{(d-1)}{\k_D^2l_D}-\frac{(d-2)}{2l_D}\f^I\f^I+
C_{IJK}\f^I\f^J\f^K+\co(\f^4),
\ee
which have the same quadratic term as $W_o(\f)$ and can therefore be
used as the counterterm. They only differ from $W_o(\f)$ at cubic order,
which corresponds to a finite counterterm. In principle, one is perfectly
allowed to use any of these fake superpotentials as the counterterm,
corresponding to a different renormalization scheme. However, since
the counterterm is valid, and the {\em same}, for any solution
of a given bulk action, there is a unique counterterm which ensures
that the action vanishes for the supersymmetric domain wall solution
defined by $W_o(\f)$. Choosing any other counterterm would simply
result in a non-supersymmetric renormalization scheme. Choosing
the supersymmetric renormalization scheme, therefore, the renormalized
on-shell action is given by
\be\label{ren-action}
S_{\rm ren}^B=\int d^dx \sqrt{\g_B}(W(\f_B)-W_o(\f_B)).
\ee

The analysis so far is independent of the dimension chosen for the
dual operator. However, we will now see that, depending on such a choice,
this renormalized action has different interpretations in the dual theory.

\subsection{$\D=\D_+$}

Consider first the more familiar case where the dimension, $\D$, of the dual
operators $\co^I_\D$ is taken to be $\D_+$. The leading asymptotic term
$\f_-$ in (\ref{generic-scalar}) corresponds then to the source of the dual
operator, since $\D_-=d-\D_+$. In this case the generating functional of
correlation functions is the renormalized on-shell action, which, evaluated on
the background domain wall solution, takes the form (\ref{ren-action}). Using
the Hamiltonian version of holographic renormalization, we find that the VEV
of the dual stress tensor is related to the extrinsic curvature of the
domain wall metric by \cite{Papadimitriou:2004rz}
\be
K^i_j=\dot{A}\d^i_j=-\frac{\k^2}{d-1}W(\f)\d^i_j.
\ee
In particular, the renormalized expectation value of the stress tensor
is given by
\be
\langle T^i_j\rangle_{\rm ren.}=-\frac{1}{\k^2}\left(K\sub{d}^i_j
-K\sub{d}\d^i_j\right)=-(W(\f)-U(\f))\d^i_j.
\ee
 Moreover, the renormalized VEV of the scalar operators is
\be\label{delta_plus_vev}
\langle \co^I_{\D_+} \rangle_{\rm ren.}=\frac{\pa}{\pa\f^I}(W(\f)-U(\f)).
\ee

The value of these one-point functions depends on the form of $W(\f)$, and
in particular, on the quadratic one, but possibly on higher order terms
as well. To be concrete, let us return to the case $d=3$ and the
$SL(8,\mathbb{R})/SO(8)$ scalars. Recall that in this case the most
general fake superpotential $W(\f)$ has two possible
forms. First, there is a continuous family of fake superpotentials
whose quadratic term is the same as that of $W_o(\f)$, but have
arbitrary cubic term. The Taylor expansion of these fake superpotentials takes
the form (\ref{W_Taylor_special}). Evaluating the one-point functions in this
case gives
\be
\langle T^i_j\rangle_+=-(C_{IJK}-C_{oIJK})\f_B^I\f_B^J\f_B^K\d^i_j,\qquad
\langle \co^I_{\D_+}\rangle=3(C_{IJK}-C_{oIJK})\f_B^J\f_B^K,
\ee
where $C_{oIJK}=\frac{\sqrt{2}\k}{96l}\sum_{i=1}^8b_{iI}b_{iJ}b_{iK}$
is the cubic coefficient of $W_o(\f)$ and the subscript $+$ is a reminder that
the VEV is taken in the theory where the operators dual to the scalar
fields have dimension $\D_+$.  It is understood that these are the
renormalized VEVs. There is also a second class of fake
superpotentials which have vanishing cubic term, but whose quadratic term can
be different from that of $W_o(\f)$. Namely, the diagonal elements of the
matrix multiplying the quadratic term of $W(\f)$ can be either $-\D_+/2l$ or
$-\D_-/2l$. Depending on which of these two values the $I$-th component takes,
in this case the one-point functions are given by
\be
\langle T^i_j\rangle_+=C_{oIJK}\f_B^I\f_B^J\f_B^K\d^i_j,\qquad
\langle \co^I_{\D_+}\rangle=\left\{\begin{matrix}
&-3C_{oIJK}\f_B^J\f_B^K,& & -\D_-/2l,\\
&-3C_{oIJK}\f_B^J\f_B^K&+\frac1l (d-2\D_+)\f_B^I, & -\D_+/2l.
\end{matrix}\right.
\ee
Note that the quadratic term of $W(\f)$ does not contribute to
the VEV of the stress tensor since $2\D_+>d$. Moreover, in the terms
involving $C_{oIJK}$ only the scalars for which the diagonal matrix
multiplying the quadratic term of $W(\f)$ has values $-\D_-/2l$ contribute,
since $3\D_-=d$ and $2\D_-=\D_+$. The components involving scalars
with a quadratic term $-\D_+/2l$ in the fake superpotential do not contribute
to these terms. We should emphasize that although we do not in general know
the full fake superpotentials, the one-point functions we have calculated
are in fact {\em exact}, since they only depend on the quadratic and cubic
terms of the fake superpotential. As a check, one can easily verify that in
both cases, the Ward identity
\be\label{Ward+}
\langle T^i_i\rangle_+=-\sum_I(d-\D_+)\f_B^I
\langle \co^I_{\D_+}\rangle,
\ee
is satisfied. Finally, for future reference, let us give more explicitly the
VEVs for the one-scalar solutions of Section \ref{one-scalar}. Namely,
\be
\begin{tabular}{lll}
$\langle T^i_j\rangle_+=\frac{\k}{l}\frac{16(\a-\a_o)}
{(2k(8-k))^{3/2}}\f_B^3\d^i_j$,
&$\langle \co_{\D_+}\rangle=
-\frac{3\k}{l}\frac{16(\a-\a_o)}{(2k(8-k))^{3/2}}\f_B^2$,& {\rm for}\;
$W(\f;\a)$,\\\\
$\langle T^i_j\rangle_+=0$,&
$\langle \co_{\D_+}\rangle=-\frac1l\f_B$, &{\rm for}
\;$\widetilde{W}_o(\f)$,
\end{tabular}
\ee
and recall that $\a_o=-(8-k)(k-4)k/24$.

The VEVs we have just computed show that if one associates
the $SL(8,\mathbb{R})/SO(8)$ scalars with operators of dimension $\D_+=2$,
the supersymmetric domain walls corresponding to $W_o(\f)$ have zero
VEVs and hence describe a deformation of the CFT Lagrangian. The
non-supersymmetric domain walls corresponding to $\widetilde{W}_o(\f)$,
however, describe a non-conformal and non-supersymmetric vacuum.
Moreover, the continuous family of domain walls defined by $W(\f;\a)$,
with $\a\neq\a_o$, gives VEVs to both the stress tensor and the scalar
operators, but these are {\em non-linear} in the scalar source. These
VEVs are on top of the deformation corresponding to $W_o(\f)$ and they 
break supersymmetry.

\subsection{$\D=\D_-$}

Consider the case where the dimension $\D$ of the dual operator is
$\D_-$. Of course, the leading asymptotic behavior of the scalar field
(\ref{generic-scalar}) is still the term involving $\f_-$, but now
it cannot be identified with the source of the dual operator since it has
the wrong asymptotic behavior for being the source. This means that the
renormalized action (\ref{ren-action}) cannot be the generating functional of
correlation functions of the operator $\co_{\D_-}$. So a more careful analysis
is required in this case.

As we have already pointed out, the evaluation of the renormalized action
by adding covariant counterterms is {\em not} affected by the question of
whether the dual operator has dimension $\D_+$ or $\D_-$.
The only difference arises in the identification of the functional
that generates the corresponding correlation functions of the
dual operator. In any case, therefore, following the standard procedure,
we need to evaluate the renormalized on-shell action, which we now call
$I[\f_-]$. This is a functional of $\f_-$ - {\em independently} of which
choice for the dimension of the dual operator is made - since the
supergravity equations of motion with Dirichlet boundary conditions
express $\f_+$ as a functional of $\f_-$. If the dimension of the
dual operator is $\D_+$, as we have seen $\f_-$ corresponds to the source
of the operator and, hence, $I[\f_-]$ can be identified with the
generating functional of connected correlators of $\co_{\D_+}$.
If the dimension of the dual operator is $\D_-$, however, this identification
cannot be made since, the still arbitrary function $\f_-$, does not correspond
to the source of $\co_{\D_-}$. In \cite{Klebanov:1999tb} it was suggested that
in this case the correct generating functional is obtained from $I[\f_-]$
by a Legendre transformation as
\be\label{legendre1}
L[\bar{\f}_-,\f_-]=I[\f_-]+\int d^dx\sqrt{g\sub{0}}\bar{\f}_-(x)\f_-(x),
\ee
where $g\sub{0}_{ij}$ is the boundary metric. Extremizing $L[\bar{\f}_-,\f_-]$
with respect to $\f_-$, gives
\be\label{legendre2}
\bar{I}[\bar{\f}_-]\equiv L[\bar{\f}_-,\f_-^*(\bar{\f}_-)],
\ee
where $\f_-^*(\bar{\f}_-)$ is the solution to
\be\label{legendre-constr}
\left.\frac{\d L[\bar{\f}_-,\f_-]}{\d \f_-}\right|_{\f_-^*}=
\left.\frac{\d I[\f_-]}{\d \f_-}\right|_{\f_-^*}+ \bar{\f}_-(x)=
\langle \co_{\D_+}\rangle_{\f_-=\f_-^*}+\bar{\f}_-(x)=0.
\ee
$\bar{I}[\bar{\f}_-]$ is now identified with the generating functional of
connected correlation functions of the operator $\co_{\D_-}$ and
$\bar{\f}_-(x)$ is identified with the source of
$\co_{\D_-}$.\footnote{Note that $\bar{\f}_-(x)\sim \f_+(x)$ up to some
numerical factor. See e.g. \cite{lectures}.} In particular, the exact
and renormalized one-point function of $\co_{\D_-}$ in the presence of a
source is
\be\label{delta-minus-vev}
\langle\co_{\D_-}\rangle_{\bar{\f}_-}\equiv \frac{\d \bar{I}[\bar{\f}_-]}
{\d \bar{\f}_-}=\f_-^*(\bar{\f}_-),
\ee
which is simply the solution to (\ref{legendre-constr}). The last
two equations tell us that the one-point function of the operator $\co_{\D_-}$
is proportional to the source, $\f_-$, of the operator $\co_{\D_+}$ and vice
versa. We can now check that an analogue of the Ward identity (\ref{Ward+})
holds for this case too. First we note that the stress tensors corresponding
to the two generating functionals $I$ and $\bar{I}$ are related by
\be\label{stress-tensor-relation}
\langle T_{ij}\rangle_{\bar{\f}_-}\equiv \frac{2}{\sqrt{g\sub{0}}}
\frac{\d \bar{I}[\bar{\f}_-,g\sub{0}]}{\d g\sub{0}^{ij}}=\langle T_{ij}
\rangle_{\f_-}- g\sub{0}_{ij}\bar{\f}_{-I}\f_-^I.
\ee
Using now the Ward identity (\ref{Ward+}) for $\langle T^i_i\rangle_{\f_-}$
together with the relations $\langle \co_{\D_+}^I\rangle=
-\bar{\f}_{-I}$ and $\langle \co_{\D_-}^I\rangle=\f_-^I$, we
obtain
\be\label{Ward-}
\langle T^i_i\rangle_{ \bar{\f}_-}=-\sum_I(d-\D_-)\bar{\f}_{-I}\langle
\co_{\D_-}^I\rangle,
\ee
where we have used $\D_-+\D_+=d$.

We can now evaluate the one-point function of the operators $\co_{\D_-}^I$
in the background domain wall solutions very easily. Starting from the
renormalized on-shell action (\ref{ren-action}), we can immediately
evaluate the one-point functions by solving (\ref{legendre-constr}), which
in this case reads:
\be
\frac{\pa}{\pa \f^I}(W(\f)-W_o(\f))+\bar{\f}_{-I}=0.
\ee
Comparing this to (\ref{delta_plus_vev}), we see that the
source $\bar{\f}_-$ of the operator $\co_{\D_-}$ is proportional to the
VEV of the operator $\co_{\D_+}$. It follows that asymptotically
$\bar{\f}_-=\co(e^{-\D_+r/l})=\co(e^{-(d-\D_-)r/l})$, as is required for the
source of an operator of dimension $\D_-$. This equation can be used to
determine $\f_B(\bar{\f}_{-B})$, but since the domain wall solution is given
in terms of $\f_B^I$ and not $\bar{\f}_{-B}^I$, we can evaluate the VEVs in
terms of $\f_B^I$, instead of $\bar{\f}_{-B}^I$. From (\ref{delta-minus-vev})
we get, depending on the coefficient of the quadratic term in $W(\f)$,
\be
\langle\co_{\D_-}^I\rangle=\left\{\begin{matrix}
\f_B^I, & -\D_-/2l, \\ 0, & -\D_+/2l.
\end{matrix}\right.
\ee
In this case, therefore, the VEVs are much simpler and completely
independent of the cubic term in $W(\f)$. Moreover, from
(\ref{stress-tensor-relation}) we find
\be\label{st}
\langle T^i_j\rangle_-=2(C_{IJK}-C_{oIJK})\f_B^I\f_B^J\f_B^K\d^i_j,
\ee
for the continuous family of fake superpotentials which has a coefficient
$-\D_-/2l$ for the quadratic term of all scalar fields, while
\be
\langle T^i_j\rangle_-=-2C_{oIJK}\f_B^I\f_B^J\f_B^K\d^i_j,
\ee
for the superpotentials that have a vanishing cubic term but any combination
of $-\D_\pm/2l$ for the quadratic term. Again, only the scalar fields with
$-\D_-/2l$ in the quadratic term contribute to the last expression.

We see that the role of $W_o(\f)$ and $\widetilde{W}_o(\f)$ have been
interchanged now compared to the case where the dual operators have dimension
$\D_+$. Namely, the supersymmetric domain walls now describe a non-conformal
but supersymmetric vacuum, which has been identified with the
Coulomb branch of the dual CFT \cite{Kraus:1998hv}, while the
non-supersymmetric domain walls corresponding to $\widetilde{W}_o(\f)$
describe a (single-trace) deformation of the CFT Lagrangian. The domain walls 
described by the continuous family of fake superpotentials $W(\f;\a)$ 
correspond to a line of marginal triple-trace deformations of the Coulomb 
branch. We stress that this does {\em not} mean that the theory has
a flat direction. At the supersymmetric point, i.e. the Coulomb branch,
the scalar operator can have an arbitrary VEV. If the marginal triple-trace
deformation is turned on, it produces a potential for the VEV of 
the scalar operator forcing it to zero. Nevertheless, it {\em is} possible
to give an arbitrary VEV to this operator in this case too, provided we 
{\em simultaneously} turn on a source for the single-trace operator,
proportional to the marginal triple-trace deformation parameter.
This single-trace deformation breaks conformal invariance 
{\em explicitly}, which justifies the fact that the trace of the stress 
tensor in (\ref{st}) is non-zero. Spontaneous breaking of the conformal
symmetry {\em only} occurs at the supersymmetric point corresponding
to no deformation. These {\em combined} marginal triple-trace and
induced single-trace deformations allowing for an arbitrary VEV
is precisely what is described by the non-supersymmetric domain walls
corresponding to $W(\f;\a)$.

\section{Holographic two-point functions}
\label{2pt-functions}

To further understand the RG flows described by the domain walls we have
discussed, we now turn to the computation of the holographic two-point
functions. However, even for the supersymmetric domain walls with a single
scalar field turned on, the linearized bulk equations of motion that we need
to solve cannot always be solved analytically. We will therefore focus on a
single scalar field, and in particular on the case $k=4$ for which the
two-point functions corresponding to the supersymmetric background can be
computed exactly. Unfortunately for $k=4$ we do not have the full
non-perturbative fake superpotentials $W(\f;\a)$ or $\widetilde{W}_o(\f)$
as for $k=6$, but we do have $W(\f;\a)$ perturbatively in $\a-\a_o$ and we can
therefore compute the two-point functions for the corresponding
non-supersymmetric backgrounds perturbatively in the parameter $\a-\a_o$.
Of course, this will provide us with no information on the domain wall defined
by $\widetilde{W}_o(\f)$, however. To evaluate these two-point functions we
will follow the approach suggested in \cite{Papadimitriou:2004ap}, where the
relevant counterterms can be evaluated directly from the linearized equations,
without the need for the computation of the full non-linear set of counterterms
required in general for the gravity-scalar action. Indeed, as we will
see, the counterterms that we will need are almost trivial. For earlier work
on the holographic computation of correlation functions see
\cite{Freedman:1998tz,DeWolfe:2000xi,Muck:2001cy, Bianchi:2001de,
Bianchi:2003ug}.

To calculate the sought after two-point functions, we need to linearize the
bulk equations of motion around the domain wall background (\ref{poincare}).
To this end, we write the bulk metric in the form
\be
ds^2=dr^2+\g_{ij}(r,x)dx^idx^j,
\ee
and consider linear fluctuations
\be
\g_{ij}=\g_{Bij}(r)+h_{ij}(r,x)=e^{2A(r)}\d_{ij}+h_{ij}(r,x),\qquad
\f=\f_B(r)+\vf(r,x).
\ee
The extrinsic curvature, $K_{ij}=\frac12\dot{\g}_{ij}$, then becomes
\be
K^i_j=\dot{A}\d^i_j+\frac12\dot{S}^i_j,
\ee
where $S^i_j\equiv \g^{ik}_Bh_{kj}$. Next we decompose $S^i_j$ into irreducible
components as
\be
S^i_j=e^i_j+\partial^i\e_j+\partial_j\e^i+\frac{d}{d-1}
\left(\frac{1}{d}\d^i_j-\frac{\partial^i\partial_j}{\square_B}\right)f+
\frac{\partial^i\partial_j}{\square_B}S,
\ee
where $\partial_i e^i_j=e^i_i=\partial_i\e^i=0$,
$\square_B=e^{-2A}\square=e^{-2A}\d^{ij}\partial_i\partial_j$, and indices
are raised with the inverse background metric $e^{-2A}\d^{ij}$. Conversely,
the projection operators
\be
\P^i\phantom{}_k\phantom{}^l\phantom{}_j=\frac12\left(\p^i_k\p^l_j+
\p^{il}\p_{kj}-\frac{2}{d-1}\p^i_j\p^l_k\right),
\ee
and
\be
\p^i_j=\d^i_j-\frac{\partial^i\partial_j}{\square_B},
\ee
allow one to uniquely express each of the irreducible components in terms of
$S^i_j$ as
\be
e^i_j=\P^i\phantom{}_k\phantom{}^l\phantom{}_j S^k_l,\,\,\,\,
\e_i=\p^l_i\frac{\partial_k}{\square_B}S^k_l,\,\,\,\,f=\p^l_k S^k_l,\,\,\,\,
S=\d^l_k S^k_l.
\ee

The linearized equations for these modes are \cite{Papadimitriou:2004ap}
\bea\label{linearized_eqs}
&&\left(\partial_r^2+d\dot{A}\partial_r+e^{-2A}\square\right) e^i_j=0,\NO\\
&&\left(\partial_r^2+[d\dot{A}+2W\partial^2_\f\log W]\partial_r+e^{-2A}
\square\right)\om=0,\NO\\
&&\dot{f}=-2\k^2\dot{\f}_B\vf,\NO\\
&&\dot{S}=\frac{1}{(d-1)\dot{A}}\left[-e^{-2A}\square f+2\k^2\left(\dot{\f}_B
\dot{\vf}-V'(\f_B)\vf\right)\right],
\eea
where
\be
\om\equiv \frac{W}{W'}\vf+\frac{1}{2\k^2}f.
\ee
Note that in writing the linearized equations in this form  we have used the
diffeomorphism invariance in the transverse space to set $\e_i\equiv 0$.
The exact, unrenormalized, one-point
functions in the presence of linear sources are given by the canonical
momenta $\dot{e}^i_j,\;\dot{\om}$ etc. The last two equations give
immediately the momenta dual to $f$ and $S$. To determine
the momenta for $e^i_j$ and $\om$ we note that, to linear order, we must
have \cite{Papadimitriou:2004ap}
\be
\dot{e}^i_j=E(A,\f_B)e^i_j,\qquad\dot{\om}=\Om(A,\f_B)\om.
\ee
Inserting these relations into the first two equations in
(\ref{linearized_eqs}), we obtain two first order equations for $E$ and $\Om$
\bea\label{firstorderequations}
&&\dot{E}+E^2+d\dot{A}E-e^{-2A}p^2=0,\NO \\
&&\dot{\Om}+\Om^2+[d\dot{A}+2W\partial^2_\f\log W]\Om-e^{-2A}p^2=0,
\eea
where $p$ denotes the transverse space momentum. All canonical
momenta can now be easily expressed in terms of $E$ and $\Om$. From
(\ref{linearized_eqs}) we deduce
\bea\label{linearized_mom}
&&\dot{e}^i_j=Ee^i_j,\NO\\
&&\dot{f}=-2\k^2W'\vf,\NO\\
&&\dot{\vf}=(W''+\Om)\vf+\frac{1}{2\k^2}\frac{W'}{W}\Om f,\NO\\
&&\dot{S}=-\frac{1}{\k^2}\left[\left(\frac{W'}{W}\right)^2\Om+
\frac{e^{-2A}}{W}p^2\right]f-2\frac{W'}{W}\left(\Om+\frac{d\k^2}{d-1}W
\right)\vf.
\eea

As we have pointed out, these are the {\em unrenormalized} momenta. To
get the renormalized momenta we need to add appropriate counterterms.
In the Hamiltonian formulation of holographic renormalization the
counterterms for the canonical momenta are simply computed by expanding
the latter in eigenfunctions of the dilatation operator
\cite{Papadimitriou:2004ap}
\be
\d_D\equiv \int d^dx 2\g_{ij}\frac{\d}{\d\g_{ij}}+(\D-d)\int d^dx \f
\frac{\d}{\d\f}.
\ee
As long as we are interested in the canonical momenta to linear order in
the sources, which is sufficient for computing the two-point functions,
expanding the expressions (\ref{linearized_mom}) in
eigenfunctions of the dilatation operator is particularly simple, since
we only need to expand the coefficients of the linear fluctuations. These
are only functions of the background fields $A$ and $\f_B$ and so
the dilatation operator simplifies to
\be
\d_D=\partial_A+(\D-d)\f_B\partial_{\f_B}.
\ee
It is important to keep in mind that, although the covariant counterterms
for the canonical momenta are computed directly by expanding the
canonical momenta in eigenfunctions of the dilatation operator, the same
result would be obtained by first computing the covariant counterterms
for the on-shell action and then deriving the renormalized momenta from
the renormalized on-shell action. Indeed, the ability to compute
the renormalized momenta, which are linear in the fluctuations, without first
having to compute the renormalized on-shell action, which is quadratic in
the fluctuations, is one of the advantages of the Hamiltonian approach.
One must, however, ensure that the counterterms for the canonical momenta
correspond to a given renormalization scheme, which is usually determined
by fixing the value of the renormalized on-shell action on a given
background. Since the only non-vanishing contribution to the counterterms,
when evaluated on the background domain wall, is the function $U(\f)$
(see Section \ref{1-pt-fns}), a given renormalization scheme is defined
by a choice of $U(\f)$. It follows that the renormalized momenta will
automatically be compatible with the chosen renormalization scheme once
the contribution of $U(\f)$ to the counterterms of the canonical momenta
has been taken into account. As we have seen in Section \ref{1-pt-fns},
for the domain walls we are interested in here we can take $U(\f)$
to be the superpotential $W_o(\f)$, corresponding to a supersymmetric
renormalization scheme. The corresponding counterterm
\be
-\int d^dx\sqrt{\g}W_o(\f_B+\vf),
\ee
leads to the following contributions to the canonical momenta:
\bea
&\dot{S}^i_j: & \frac{2\k^2}{(d-1)}W_o'(\f_B)\vf\d^i_j\NO\\
&\dot{\vf}: & -W_o''(\f_B)\vf.
\eea
Adding these contributions to the canonical momenta (\ref{linearized_mom})
we obtain
\bea\label{linearized_mom_half_ren}
&&\dot{e}^i_j=Ee^i_j,\NO\\
&&\dot{f}=-2\k^2(W'-W_o')\vf,\NO\\
&&\dot{\vf}=(W''-W_o''+\Om)\vf+\frac{1}{2\k^2}\frac{W'}{W}\Om f,\NO\\
&&\dot{S}=-\frac{1}{\k^2}\left[\left(\frac{W'}{W}\right)^2\Om+
\frac{e^{-2A}}{W}p^2\right]f-2\left(\frac{W'}{W}\Om+\frac{d\k^2}{d-1}(W'-W_o')
\right)\vf.
\eea
These are not yet the renormalized momenta, but it is now guaranteed that by
expanding these canonical momenta in eigenfunctions of the dilatation
operator and keeping the terms of weight $d=3$ for $\dot{e}^i_j,\;\dot{f}$
and $\dot{S}$, and of weight $\D_+=2$ for $\dot{\vf}$, we obtain the
renormalized momenta that correspond to the supersymmetric scheme defined
by $W_o$.

Since $W(\f_B)$ and $W_o(\f_B)$ are known functions of the background
fields, the only non-trivial step in computing the renormalized canonical
momenta is determining the expansion of $E$ and $\Om$ in eigenfunctions of the
dilatation operator. This can be easily done by using the
equations (\ref{firstorderequations}). One expands the radial derivative as
\be
\partial_r=\dot{A}\partial_A+\dot{\f}_B\partial_{\f_B}=
-\frac{\k^2}{d-1}W(\f_B)\partial_A+W'(\f_B)\partial_{\f_B}\sim \d_D+\cdots,
\ee
as well as the functions $E$ and $\Om$\footnote{In general one would have
to include logarithmic terms in these expansions, but in our case we do
not need them since the boundary is three-dimensional. See
\cite{Papadimitriou:2004ap} for the general case.}
\bea\label{expansions}
&&E=E\sub{1}+\cdots+E\sub{d}+\cdots,\NO \\
&&\Om=\Om\sub{0}+\cdots+\Om\sub{2\D-d}
+\cdots,
\eea
and inserts these expansions in equations (\ref{firstorderequations}).
Collecting terms of the same dilatation weight then determines
all terms in the expansions (\ref{expansions}), except for the
coefficients $E\sub{d}$ and $\Om\sub{2\D-d}$. These terms contain
all dynamical information about the two-point functions and can only
be determined by solving exactly the first two equations in
(\ref{linearized_eqs}) or, equivalently, equations (\ref{firstorderequations}).
There is, however, an important technical difference between solving the
former or the latter. By solving the linear second order
equations in (\ref{linearized_eqs}), one obtains two linearly independent
solutions, namely the `normalizable' and `non-normalizable'
modes. Generically, an arbitrary linear combination of these solutions
will have a singularity somewhere in the interior of the asymptotically AdS
space. However, there is usually a unique linear combination which leads
to a non-singular solution. This requirement determines the coefficient
of the normalizable mode in terms of the non-normalizable mode, which
should be arbitrary since it corresponds to the source of the dual operator.
If one instead solves the first order equations (\ref{firstorderequations})
only one integration constant appears instead of two,  which simply
reflects the fact that the overall normalization of the linearized
solutions of (\ref{linearized_eqs}) has been factored out from $E$
and $\Om$. The integration constant in $E$ and $\Om$ can therefore be
understood as the ratio of the normalizable and non-normalizable modes of
the solutions to the second order equations (\ref{linearized_eqs}).
However, it is not always possible to determine this integration
constant by the requirement that the exact solutions for $E$ and $\Om$
are non-singular. This is because it is possible that $E$ and $\Om$
are non-singular, even though the corresponding solutions of the
second order equations (\ref{linearized_eqs}) are singular. Practically,
therefore, to obtain the correct exact solution for $E$ and $\Om$, one
should first solve the corresponding second order equations
(\ref{linearized_eqs}), demand that they are non-singular, and then
deduce the corresponding $E$ and $\Om$. Equations (\ref{firstorderequations})
are still essential, however, for determining the covariant counterterms
for $E$ and $\Om$.

We compute $E\sub{3}$ and $\Om\sub{1}$ explicitly in Appendix
\ref{2pt-computation} for the $k=4$ one-scalar domain wall. The
result is given in (\ref{dynamic}). Given these quantities one
can now determine the one-point functions with linearized sources
and, consequently, the exact two-point functions. Expanding
the canonical momenta (\ref{linearized_mom_half_ren}) one easily deduces
that the renormalized one-point functions are given by
\bea\label{exact-1pt-fns+}
&&\langle T^i_j\rangle_+ =\frac{\k\a}{8\sqrt{2}l}\f_B^3\d^i_j
-\frac{1}{2\k^2}E\sub{3}e\sub{0}^i_j-\frac18 \f_B^2\p^i_j\Om\sub{1}f\sub{0}
+\left(\frac{3\k\a}{8\sqrt{2}l}\f_B^2\d^i_j-\frac12\f_B\p^i_j\Om\sub{1}\right)
\vf\sub{0},\NO\\
&&\langle \co_{\D_+}\rangle =-\frac{3\k\a}{8\sqrt{2}l}\f_B^2+\frac14\f_B
\Om\sub{1}f\sub{0}+\left(-\frac{3\k\a}{4\sqrt{2}l}\f_B+\Om\sub{1}\right)
\vf\sub{0}.
\eea
Again, one should keep in mind that these are the one-point functions
when the dual scalar operators are taken to have dimension $\D_+$. We
will consider the case $\D_-$ below. It is reassuring that these one-point
functions satisfy the Ward identity (\ref{Ward+}) as they should.
Differentiating with respect to the linear sources one finally obtains the
two-point functions
\bea
&&\langle T^i_jT^k_l\rangle =-\frac{1}{\k^2}\P^i\phantom{}_l\phantom{}^k
\phantom{}_j E\sub{3}-\frac14\f_B^2\p^i_j\p^k_l\Om\sub{1},\NO\\
&&\langle T^i_j\co_{\D_+}\rangle =-\frac{3\k\a}{8\sqrt{2}l}\f_B^2\d^i_j+
\frac12\f_B\p^i_j\Om\sub{1},\NO\\
&&\langle\co_{\D_+}\co_{\D_+}\rangle =\frac{3\k\a}{4\sqrt{2}l}\f_B-\Om\sub{1}.
\eea

The two-point functions for the case where the dual scalar operators have
dimension $\D_-$ can also be deduced from the one-point functions
(\ref{exact-1pt-fns+}). As we have seen in the previous section, the source
dual to $\co_{\D_-}$ is given by the VEV of $\co_{\D_+}$ as
\be
\bar{\f}=\bar{\f}_{B}+\bar{\vf}=-\langle \co_{\D_+}\rangle,
\ee
from which we infer
\bea
\vf & = &\left(\frac{3\k\a}{4\sqrt{2}l}\f_B-\Om\sub{1}\right)^{-1}
\left(\bar{\vf}+\frac 14\f_B\Om\sub{1}f\sub{0}\right)\NO\\
& = &- \Om\sub{1}^{-1}
\left(1+\frac{3\k\a}{4\sqrt{2}l}\f_B\Om\sub{1}^{-1}+\co(\a^2)\right)
\left(\bar{\vf}+\frac 14\f_B\Om\sub{1}f\sub{0}\right).
\eea
It follows that
\be
\langle \co_{\D_-}\rangle =(\f_B+\vf)=
\f_B-\left(1+\frac{3\k\a}{4\sqrt{2}l}\f_B\Om\sub{1}^{-1}
+\co(\a^2)\right)\left(\Om\sub{1}^{-1}\bar{\vf}+\frac14\f_Bf\sub{0}
\right).
\ee
Moreover, from (\ref{stress-tensor-relation}) we obtain
\bea
\langle T^i_j\rangle_- & = &\langle T^i_j\rangle_+-\bar{\f}\langle \co_{\D_-}
\rangle \d^i_j\NO\\
& = &-\frac{\k\a}{4\sqrt{2}l}\f_B^3\d^i_j-\frac{1}{2\k^2}E\sub{3}e\sub{0}^i_j
-\f_B\left(\d^i_j-\frac12\p^i_j\left(1+\frac{3\k\a}{4\sqrt{2}l}\f_B
\Om\sub{1}^{-1}+\co(\a^2)\right)\right)\bar{\vf}\NO\\
&& +\frac{3\k\a}{32\sqrt{2}l}\f_B^3\p^i_jf\sub{0}.
\eea
Once again, these one-point functions satisfy the Ward identity (\ref{Ward-})
as required. Differentiating with respect to the sources we now obtain
the two-point functions
\bea
&&\langle T^i_jT^k_l\rangle =-\frac{1}{\k^2}\P^i\phantom{}_l\phantom{}^k
\phantom{}_j E\sub{3}+\frac{3\k\a}{16\sqrt{2}l}\f_B^3\p^i_j\p^k_l,\NO\\
&&\langle T^i_j\co_{\D_-}\rangle = \f_B\left(\d^i_j-\frac12\p^i_j
\left(1+\frac{3\k\a}{4\sqrt{2}l}\f_B\Om\sub{1}^{-1}+\co(\a^2)\right)\right),
\NO\\
&&\langle\co_{\D_-}\co_{\D_-}\rangle=\Om\sub{1}^{-1}
\left(1+\frac{3\k\a}{4\sqrt{2}l}\f_B\Om\sub{1}^{-1}+\co(\a^2)\right).
\eea

In order to discuss the physics of these two-point functions, it
is useful to reinstate the dependence on the $M2$-brane distribution
parameters, which we have so far suppressed because this is different
for different domain walls. For the case $k=4$ for which we have 
computed the two-point functions, the $M2$-branes are distributed
on an $S^3$ of radius $l_1$. By uplifting our domain wall solution to
eleven-dimensions we find that the scalar VEV is as expected proportional
to the radius of the $M2$-branes distribution, $\f_B=\sqrt{2}l_1/\k l$, 
while the momenta, $\tilde{p}^i$, on the world-volume of the $M2$-branes
are related to the momenta above by $p^i=(l/l_1)\tilde{p}^i$. With these
relations and the result (\ref{dynamic}) for $\Om\sub{1}$ we can
write the scalar two-point function as
\be
\langle\co_{\D_-}\co_{\D_-}\rangle=\frac{2}{\tilde{p}^2}
\sqrt{\tilde{p}^2+\frac{4l_1^2}{l^4}}\left(1-\a l_1
\frac{(9l^8\tilde{p}^4+32l_1^2l^4\tilde{p}^2+16l_1^4)}{2l^{10}\tilde{p}^2
(\tilde{p}^2+4l_1^2/l^4)^{3/2}}+\co(\a^2)\right).
\ee 

We can now extract the physics. First, for $\a=0$, there is a massless 
Goldstone
pole corresponding to the spontaneously broken scale invariance. Moreover,
there is a continuous spectrum of states corresponding to the branch cut
$(4l_1^2/l^4,+\infty)$ on the complex {\em Lorentzian} $\tilde{p}^2_L=-
\tilde{p}^2$ plane. Note that the threshold $M^2=4l_1^2/l^4$ agrees
precisely with that found in \cite{Bakas:1999fa} by different means.
Moreover, in the limit of vanishing VEV, $l_1\to 0$, we restore the two-point
function imposed by conformal invariance for an operator of dimension 1
in three dimensions. Note in particular that in this limit the deformation
parameter does not modify the two-point function, at least to the 
order we have computed it. This suggests that the $\co_{\D_-}$ does
not acquire an anomalous dimension when the marginal deformation is turned on,
again at least to the order in $\a$ we have computed it and in the large-$N$ 
limit, for which the supergravity approximation holds.  

\section{The fake superpotential as a quantum effective potential
and multi-trace deformations}
\label{eff-pot}

We will now argue that, under certain circumstances, the fake superpotential 
that defines a given domain wall has a direct physical interpretation in the
dual field theory as a quantum effective potential describing a marginal
multi-trace deformation. As we will see, this interpretation requires that 
the bulk scalar fields admit two quantizations, as is the case for the 
$SL(N,\mathbb{R})/SO(N)$ scalars that we have been discussing. In this case 
the on-shell supergravity action plays two roles. More specifically, since 
$I[\f_-]$ and $\bar{I}[\bar{\f}_-]$ are the Legendre transform of each other 
and  $\langle \co_{\D_+}\rangle=- \bar{\f}_{-}$ and $\langle 
\co_{\D_-}\rangle=\f_-$, it follows that
\begin{itemize}
{\em
\item $I[\f_-]$ is the generating functional of connected correlation
functions of $\co_{\D_+}$ and the quantum effective action for $\co_{\D_-}$.

\item $\bar{I}[\bar{\f}_-]$ is the generating functional of connected
correlation functions of $\co_{\D_-}$ and the quantum effective action for
$\co_{\D_+}$.
}
\end{itemize}

We have seen above that on a domain wall solution defined by the fake
superpotential $W(\f)$, the renormalized on-shell supergravity action
computed with the standard Dirichlet boundary conditions is  
\be
I=\int d^dx\sqrt{\g}(W(\f)-W_o(\f)).
\ee
We will now show that this relation implies that the freedom in the
fake superpotential, $W(\f)$, is equivalent to computing the on-shell
action with modified conformal boundary conditions and hence to 
a marginal multi-trace deformation of the boundary theory. 

Multi-trace operators in any QFT that admits a large-$N$ limit and
in the AdS/CFT correspondence are discussed in detail in Appendix 
\ref{multitrace}. In the AdS/CFT correspondence, the effect of deforming the 
CFT action by a multi-trace operator $f(\co_{\D_+})$, for the $\D_+$ 
quantization, or by $\bar{f}(\co_{\D_-})$, for the $\D_-$ quantization, can 
be summarized in equations (\ref{D+shift}) and (\ref{D-shift}) respectively.
It follows that the effect of a generic fake superpotential, $W(\f)$, can be
reproduced by computing the on-shell action with the superpotential 
$W_o(\f)$ but with boundary conditions corresponding to a deformation
\bea
\bar{f}(\f)& = & W(\f)-W_o(\f), \qquad {\rm for} \quad \D_-,\NO\\
f(\f)-\f f'(\f) & = & W(\f)-W_o(\f), \qquad {\rm for} \quad \D_+,
\eea
or equivalently
\be
f(\f)=-\f\int \frac{d\f}{\f^2}\left(W(\f)-W_o(\f)\right), \qquad 
{\rm for} \quad \D_+.
\ee
However, since the arguments of $f$ and $\bar{f}$ are the VEVs $\s$ and
$\bar{\s}$ respectively, this interpretation of the fake superpotential
is possible only when $W=W_+$ for the $\D_+$ quantization and $W=W_-$ for
the $\D_-$ quantization. In summary, then
\bea
\bar{f}(\f)& = & W_-(\f)-W_o(\f)= \co(\f^3), \quad {\rm for} \quad \D_-,\NO\\
f(\f) & = & -\f\int \frac{d\f}{\f^2}\left(W_+(\f)-W_o(\f)\right)= 
-\frac{1}{2l}(\D_--\D_+)\f^I\f^I+\co(\f^3), \quad {\rm for} \quad \D_+,
\eea
where we have used the expansion
\be
W_\pm(\f)=-\frac{(d-1)}{\k^2l}-\frac{1}{2l}\D_\pm \f^I\f^I+\co(\f^3).
\ee
But, if $W=W_+$, then $\f^2$ corresponds to an irrelevant operator since
$\D_+>d/2$. It is therefore only for the $\D_-$ quantization and for the
fake superpotential $W_-$ that an interpretation as a marginal multi-trace
deformation of the boundary theory can arise. The above discussion can
then be summarized in the following statement:
\begin{center}
\begin{tabular}{|l|}\hline
{\em The marginal multi-trace deformations of a CFT admitting a 
holographic dual and}\\ {\em corresponding to the $\D_-$ quantization 
are in one-to-one correspondence with the }\\ {\em possible fake 
superpotentials $W_-$ of the dual bulk theory.} \\ 
\hline
\end{tabular}
\end{center}

These fake superpotentials are determined by solving equation 
(\ref{potential}) as a differential equation for the fake 
superpotential. The conditions for such an interpretation of the fake 
superpotential impose strict restrictions on the dimension $d$ of the field 
theory as well as on the conformal dimension $\D_-$ of the local operator. 
First, in order to get an $n$-trace {\em marginal} operator built from
the single-trace operator $\co_{\D_-}$, we obviously need $\D_-=d/n$, where
$n>2$ is an integer. The condition $n>2$ arises because the fake 
superpotential, $W_-$, cannot describe double-trace deformations since there 
is no freedom in its quadratic term. Moreover $\D_-$ is bounded by 
$(d-2)/2\leq \D_- <d/2$. The possible solutions of these 
conditions are summarized in Table \ref{dimensions}. Note that only for 
$d=3$ is there an allowed $\D_-$ which is integer and yet
it does not saturate the unitarity bound, namely $\D_-=1$.
\TABLE{
\begin{tabular}{|c||c|c|c|c|c|}
\hline
$d$ & 2 & 3 & 4 & 5 & 6  \\
\hline
$n$ & $n\geq 3$ & 3, 4, 5, 6 &  3, 4 & 3 & 3 \\
\hline
$\D_-$ & $2/n$  & $1, 3/4,  3/5, \mathbf{1/2}$ & $ 4/3, \mathbf{1}$ & 
$5/3$ & $\mathbf{2}$\\
\hline
\end{tabular}
\caption{The possible dimensions $d$ and conformal dimensions $\D_-$
allowing for the interpretation of the fake superpotential as a
multi-trace deformation of the dual theory. The dimensions in boldface 
saturate the unitarity bound.}
\label{dimensions}}

\subsection{Triple-trace deformation of the Coulomb branch}

The field theory on the worldvolume of $N+1$ $M2$-branes is an $\cn=8$
(16 supercharges) superconformal field theory with $8(N+1)$ scalar
and $8(N+1)$ fermionic degrees of freedom. Under the $SO(8)$ R-symmetry
group, the scalars, fermions and supersymmetries transform respectively
as ${\bf 8}_v$, ${\bf 8}_c$ and ${\bf 8}_s$ (see e.g. \cite{Corrado:2001nv}).
One of the $\cn=8$ multiplets corresponds to the free theory describing the
center of mass motion, while the remaining degrees of freedom parameterize
the moduli space $(\mathbb{R}^8)^N/S_{N+1}$. This theory is believed to
arise as the infra red fixed point of $\cn=8$ supersymmetric Yang-Mills
in three dimensions, while in the abelian case it can also be obtained 
by compactifying $\cn=4$ supersymmetric Yang-Mills in four dimensions
on a circle in the limit of vanishing circle radius. For the 
non-abelian case, however, this procedure is not well understood.
In the large-$N$ limit this theory is holographically dual to 
eleven-dimensional supergravity on
$AdS_4\times S^7$, whose massless sector is described by $\cn=8$ gauged
supergravity in four dimensions. The 70 scalars parameterizing the
moduli space $E_{(7)7}/SU(8)$ of $\cn=8$ gauged supergravity are
holographically dual to BPS operators, which in the abelian case 
can be understood in terms the traceless
bilinears of the 8 scalars and 8 fermions:
\bea
\co^{IJ}=\Tr(X^IX^J)-\frac18\d^{IJ}\Tr(X^KX^K),\quad I,J,\ldots=1,\ldots,8\NO\\
\cp^{AB}=\Tr(\l^A\l^B)-\frac18\d^{AB}\Tr(\l^C\l^C),\quad A,B,\ldots=1,\ldots,8.
\eea
The 35 operators $\co^{IJ}$ have conformal dimension $\D=1$  and transform in
the $\mathbf{35}_v$ of $SO(8)$, while the 35 scalars $\cp^{AB}$ have
conformal dimension $\D=2$ and transform in the $\mathbf{35}_c$.
The 35 scalars parameterizing the $SL(8,\mathbb{R})/SO(8)$ subspace of the
scalar manifold are usually identified as dual to the dimension 1 operators
$\co^{IJ}$, in which case the supersymmetric domain wall solutions with
non-trivial $SL(8,\mathbb{R})/SO(8)$ scalars describe a uniform subsector of
the Coulomb branch of the $M2$-brane theory
\cite{Kraus:1998hv,Cvetic:1999xx, Bakas:1999fa}. Nevertheless, since
the $SL(8,\mathbb{R})/SO(8)$ scalars have a mass allowing two possible 
quantizations as we have discussed, and since in this case the $\D=1$ and 
$\D=2$ scalars both belong to the massless $\cn=8$ supermultiplet and 
transform in $SO(8)$ representations which are related by triality, it seems 
plausible that one could also identify these scalars as dual to the dimension 
2 operators $\cp^{IJ}$.\footnote{This is special to four dimensions. In five or 
seven dimensions, one can unambiguously identify the $SL(N,\mathbb{R})/SO(N)$ 
scalars from symmetries, by looking at the states of the relevant massless
supermultiplet. I would like to thank Henning Samtleben for useful comments 
on this.} In this case, as we have seen, the supersymmetric domain walls of the
$SL(8,\mathbb{R})/SO(8)$ sector correspond to deformations of the CFT
Lagrangian.

When the $SL(8,\mathbb{R})/SO(8)$ scalars are identified as dual to the
dimension 1 operators $\co^{IJ}$, the non-supersymmetric domain walls we
have found describe the deformation of the uniform sector of the
Coulomb branch, which corresponds to the supersymmetric superpotential,
by a marginal triple-trace operator completely breaking supersymmetry. Such
deformations have been discussed before in \cite{Hertog:2004dr, Hertog:2004rz,
Hertog:2004ns, Hertog:2005hu, Elitzur:2005kz}. Indeed, as we have argued above,
the fake superpotentials $W$ we have found compute the exact large-$N$
effective potential, given by
\be
V_{\rm eff.}(\f)=\bar{f}(\f)=W(\f)-W_o(\f)=(C_{IJK}-C_{oIJK})\f^I\f^J\f^K
+\co(\f^4).
\ee
The higher order terms, which vanish when the cut-off is removed, correspond
to irrelevant operators. Therefore, when the regulator is removed we are
left with the triple-trace operator $(C_{IJK}-C_{oIJK})\co^I\co^J\co^K$. This 
operator is classically marginal and remains marginal to leading order in 
$1/N$ for any finite value of the dimensionless moduli $C_{IJK}$. However, 
the limit $C_{IJK}\to\infty$, corresponding to replacing $W(\f;\a)$
by $\widetilde{W}_o(\f)$, does not commute with the cut-off removal,
presumably due to the fact that we are working in the large-$N$ limit. 
Hence, we cannot simply drop the higher order irrelevant operators 
altogether, but instead we need the full fake superpotential. The role
of these higher order terms can be understood by recalling that the
these domain walls describe the Coulomb branch of the dual CFT and therefore,
conformal invariance is spontaneously broken. This means that the 
coupling $C_{IJK}$ will run or, equivalently, the operator $\co^3$ will
renormalize. Since there are no other free parameters in the fake 
superpotential other than the coupling $C_{IJK}$, the effect of the 
irrelevant operators can be interpreted as the running of the coupling
$C_{IJK}$ or, equivalently, as the multiplicative renormalization of  $\co^3$.

Luckily we have already computed the exact fake superpotential for the
one-scalar domain wall with $k=6$, which we will now consider as an example. 
For this case the full fake superpotential was given in 
(\ref{exact-fake-super}). Using this we can now extract the exact large-$N$ 
anomalous dimensions $\g_\co$ and $\g_{\co^3}$. Starting from the kinetic 
term for the effective scalar field $\f$ away from the origin of the moduli 
space \cite{Elitzur:2005kz} ($\f\neq 0$ since there is a non-zero VEV)
\be
K[\f]=\frac{N^2}{8}\f^{-1}\pa_i\f\pa^i\f,
\ee
and writing $\f=\f_\L=\frac{\L_o}{\L}\tilde{\f}\left(\frac{\L_o}{\L}\right)$,
with $\tilde{\f}\left(\frac{\L_o}{\L}\right)=\co(\L^0)$ as $\L\to\infty$, and
identifying the UV cut-off with the AdS radial coordinate as $\L=e^{r/l}$,
we see that the multiplicative renormalization of $\co$ is given by
\be
Z_\co^{-1}(\L)=\tilde{\f}\left(\frac{\L_o}{\L}\right)/
\tilde{\f}(1).
\ee
It now follows from the first order equations (\ref{flow_eqs}) that
\be
\g_\co=\left.\L\pa_\L\log Z_\co(\L)\right|_{\L\to\infty}=-1-l\left.\frac{W'}
{\f}\right|_{\f\to 0}.
\ee
Evaluating this using the fake superpotential (\ref{exact-fake-super})
we obtain
\be
\g_\co=\left\{\begin{matrix} 0, && \a<\infty, \\
1, && \a\to\infty.
\end{matrix}\right.
\ee
The dimension of the operator $\co$ therefore jumps from $1$ for finite
$\a$ to $2$ in the $\a\to \infty$ limit. In this sense then the marginal
triple-trace deformation interpolates between the two possible quantizations
of the bulk scalar field, much like the situation described in e.g.
\cite{Witten:2001ua,Gubser:2002vv}.

Similarly, we can now evaluate the anomalous dimension $\g_{\co^3}$. Since
\be
V_{\rm eff.}=W-W_o=-\frac{2\k}{ l}\left(\frac32\right)^{3/2}\frac{(\a+1)}{27}
\f^3+\co(\f^4)\equiv -\frac{2\k}{ l}\left(\frac32\right)^{3/2}\frac{(\a+1)}{27}
Z_{\co^3}^{-1}\f^3,
\ee
we find
\be
\g_{\co^3}=\left.\L\pa_\L\log Z_{\co^3}(\L)\right|_{\L\to\infty}=
\left\{\begin{matrix} 0, && \a<\infty, \\
1, && \a\to\infty.
\end{matrix}\right.
\ee
Note that the running coupling is simply given by 
$\bar{\a}+1=Z_{\co^3}^{-1}(\a+1)$ and so $\b_\a=(\a+1)\g_{\co^3}$. 
It follows that, in agreement with the expectation in \cite{Hertog:2004rz},
the triple-trace operator $\co^3$ remains marginal for all finite values
of $\a$ and in the large-$N$ limit. In the $\a\to\infty$ limit, however,
$\co^3$ has dimension $4$ and not the naively expected dimension $6$. It
nevertheless remains an irrelevant operator in this limit.

\acknowledgments

I am grateful to Professor Jerome Gauntlett for a helpful conversation
on the significance of curvature singularities in M-theory backgrounds.
I would also like to thank Dimitry Belyaev for useful comments.

\appendix

\section{Explicit form of the domain wall metric for $W(\f;\a)$, to
first order in $\a-\a_o$ and for general $k$}
\label{X-coord}

In this appendix we give the explicit form of the domain wall backgrounds
corresponding to the fake superpotential (\ref{pert-superpotential}) to
first order in $\a-\a_o$. In order to discuss all values of $k$ at once, it
is convenient to trade the radial coordinate $r$ in (\ref{poincare}) for
the single scalar field $X$. Another advantage of this radial coordinate
is that it is directly related to the scalar field $\f$ via (\ref{X})
and so we only need to determine the domain wall metric. This can be done
by solving the flow equations (\ref{flow_eqs}), which now become
\bea
&&\dot{X}=\k^2\left(\frac{8-k}{2k}\right)X^2\pa_X W,\NO\\
&&\pa_X A=-\left(\frac{k}{8-k}\right)\frac{W}{X^2\pa_X W}.
\eea
Inserting the fake superpotential (\ref{pert-superpotential}) and integrating
these equations to first order in $(\a-\a_o)$ we find
\bea
\frac{dr}{l}& = &-\frac{8}{(8-k)}\frac{X^{\frac{k}{(8-k)}-1}dX}{(X^{\frac{8}
{(8-k)}}-1)}\left(1-\frac{3(\a-\a_o)}{2k(8-k)}X^{\frac{k}{(8-k)}-3}(X^{\frac{8}
{(8-k)}}-1)(kX^{\frac{8}{(8-k)}}+8-k)\right.\NO\\&&\left.+\co((\a-\a_o)^2)
\rule[.2cm]{0cm}{.3cm}\right),\NO\\
e^{A} & = & \frac{8}{(8-k)}\frac{X}{(X^{\frac{8}{(8-k)}}-1)}\left\{1
+\frac{(\a-\a_o)}{2k(8-k)}\left[k X^{\frac{4(k-2)}{(8-k)}}
+\frac{2k(8-k)}{(k-4)}
\left(X^{\frac{4(k-4)}{(8-k)}}-1\right)\right.\right.\NO\\
&&\left.\left.-(8-k)X^{\frac{4(k-6)}{(8-k)}}-2(k-4)\right]+\co((\a-\a_o)^2)
\right\}.
\eea
Note that for $\a=\a_o$ one recovers the supersymmetric solutions of
\cite{Cvetic:1999xx,Bakas:1999fa}.

\section{Uplifting the MTZ black hole to eleven dimensions}
\label{MTZ}

The gravity-scalar system (\ref{fake_action}) in four dimensions with a single
scalar field and the potential (\ref{potential_1}) was also considered in
\cite{Martinez:2004nb}, where a topological black hole with non-trivial
scalar hair was found. It was also pointed out in \cite{Martinez:2004nb}
that this scalar potential acquires a very simple
form in the conformal frame defined by
\be
\tilde{\psi}/3=\tanh(\psi/3),\qquad \tilde{g}_{\m\n}=\cosh^2(\psi/3)g_{\m\n},
\ee
where $\tilde{\psi}=\sqrt{\frac{3\k^2}{2}}\tilde{\f}$. In this frame the
action takes the form
\be
S=\int_{\cm}d^4x\sqrt{-\tilde{g}}\left(\frac{1}{2\k^2}\tilde{R}-\frac12
\tilde{g}^{\m\n}\pa_\m\tilde{\f}\pa_\n\tilde{\f}-\frac{1}{12}\tilde{R}
\tilde{\f}^2-\tilde{V}(\tilde{\f})\right),
\ee
where
\be
\tilde{V}(\tilde{\f})=-\frac{3}{\k^2l^2}\left(1-(\k^2/6)^2\tilde{\f}^4\right).
\ee
The scalar field is now conformally coupled to gravity and the
$\tilde{\f}^4$ potential ensures that the scalar field equations are
conformally invariant. Quite remarkably this system admits an exact
instanton solution \cite{dHPP}.

The four-dimensional black hole found in \cite{Martinez:2004nb}, which we
will refer to as the MTZ black hole, reads\footnote{We have kept the notation
of \cite{Martinez:2004nb} here, hoping this will cause no confusion. Note, in
particular, that the radial coordinate, $r$, here is not related to the radial
coordinate in the domain wall metric (\ref{poincare}), and, as usual,
$\k^2=8\p G$.}
\bea
ds_4^2 & = & \frac{r(r+2G\m)}{(r+G\m)^2}\left\{-\left(\frac{r^2}{l^2}
-\left(1+\frac{G\m}{r}\right)^2\right)dt^2
+\left(\frac{r^2}{l^2}-\left(1+\frac{G\m}{r}\right)^2\right)^{-1}dr^2
+r^2d\s^2\right\},\NO\\
\f & = & \sqrt{\frac{3}{4\p G}}\tanh^{-1}\left(\frac{G\m}{G\m+r}\right),
\eea
where $d\s^2$ is the metric on a two-dimensional compact manifold, $\S_2$, of
constant negative curvature. This means that $\S_2\cong \mathbb{H}_2/\G$,
where $\mathbb{H}_2$ is the hyperbolic plane and $\G$ is a freely acting
discrete subgroup of the isometry group $O(2,1)$. This black hole has
curvature singularities at $r=0$ and $r=-2G\m$. The range of the radial
coordinate is taken $r>0$ for  $\m>0$ and $r>-2G\m$ for $\m<0$. In either
case the curvature singularity is hidden behind a horizon located at
$r_+=\frac l2 (1+\sqrt{1+4G\m/l})$, provided
\be
\m>-\frac{1}{4G}.
\ee
Note also that $\f\geq0$ for $\m>0$ and $\f\leq0$ for $\m<0$. The mass of the
black hole is given by
\be
M=\frac{\s}{4\p}\m,
\ee
where $\s$ is the area of $\S_2$, and its Hawking temperature is
\be
T_H=\frac{1}{2\p l}\left(\frac{2r_+}{l}-1\right).
\ee

As in Section \ref{uplift}, we can uplift this black hole to eleven dimensions
using the reduced ansatz (\ref{ansatz-reduced}). In terms of the scalar field
\be
X=e^{\psi/3}=e^{\k\f/\sqrt{3}}=\left(\frac{2G\m+r}{r}\right)^{1/8},
\ee
and the quantity
\be
\tilde{\D}=X\cos^2\th+X^{-3}\sin^2\th,
\ee
the eleven-dimensional metric is
\be
d\hat{s}_{11}^2=\tilde{\D}^{2/3}ds_4^2+4l^2\tilde{\D}^{-1/3}
\left\{X^3\left[(\cos^2\th+X^{-4}\sin^2\th)d\th^2+\sin^2\th d\f_4^2\right]
+X^{-1}\cos^2\th d\Om_5^2\right\},
\ee
and the four-form field strength reads
\bea
\hat{F}\sub{4} & = & \left\{\frac{r^4(r+2G\m)^2}{l(r+G\m)^4}\left[2X^2\cos^2\th
+X^{-2}(1+2\sin^2\th)\right]dr\right.\NO\\
&&\left.+2G\m lX\left[1-\frac{r^2}{l^2}\left(1+\frac{G\m}{r}\right)^{-2}\right]
\cos\th\sin\th d\th\right\}\wedge dt\wedge \bar{\e}\sub{2},
\eea
where $\bar{\e}\sub{2}$ is the volume form on $\S_2$.

\section{Computation of the holographic two-point functions}
\label{2pt-computation}

In this appendix we give the details of the computation of the
holographic two-point functions for the domain walls defined by
the fake superpotential (\ref{pert-superpotential}), which have
been constructed explicitly in Appendix \ref{X-coord}.

We start with the supersymmetric solutions corresponding to $\a=\a_o$.
For this case the first two equations in (\ref{linearized_eqs}) read
respectively
\bea
&&\left(X^2\pa_X^2+\frac{(16-5k)X^{\frac{8}{(8-k)}}+(5k-32)}{(8-k)(X^{\frac{8}
{(8-k)}}-1)}X\pa_X-q^2X^{\frac{4(k-4)}{(8-k)}}\right)e^i_j=0,\NO\\
&&\left(X^2\pa_X^2+\frac{k(16-5k)X^{\frac{8}{(8-k)}}+(8-k)(5k-32)}{(8-k)
(kX^{\frac{8}{(8-k)}}+8-k)}X\pa_X-q^2X^{\frac{4(k-4)}{(8-k)}}\right)\om=0,
\eea
where $q^2=p^2l^2(8-k)^2/64$. We have managed to
solve these equations analytically only for the case $k=4$, and so we will
focus on this case. Having solved the supersymmetric case, we can then
obtain the solution for the non-supersymmetric fake superpotential $W(\f;\a)$
perturbatively in $\a-\a_o$. Since for $k=4$ we have $\a_o=0$, the expansion
is actually in $\a$. For $k=4$ then, to first order in $\a$, the first
two equations in (\ref{linearized_eqs}) take the form
\bea
&&\left\{\left(X^2\pa_X^2-\frac{(X+3X^{-1})}{(X-X^{-1})}X\pa_X-q^2\right)
\right.\\&&\left.+\frac{3\a}{4}\left(\rule[.2cm]{0cm}{.3cm}(X^2-X^{-2})
(X^2\pa_X^2+X\pa_X)+\frac{q^2}{3}(X^2-X^{-2}+8\ln X)\right)
+\co(\a^2)\right\}e^i_j=0,\NO\\\NO\\
&&\left\{\rule[.2cm]{0cm}{.5cm}\left(X^2\pa_X^2-\frac{(X-3X^{-1})}{(X+X^{-1})}
X\pa_X-q^2\right)\right.\\&&\left.+\frac{3\a}{4}\left((X^2-X^{-2})X^2
\pa_X^2+\left[X^2-X^{-2}+\frac43\left(\frac{(X+X^{-1})^2+2}{X+X^{-1}}
\right)^2\right]X\pa_X\right.\right.\NO\\
&&\left.\left.+\frac{q^2}{3}(X^2-X^{-2}+8\ln X)\rule[.2cm]{0cm}{.5cm}\right)
+\co(\a^2)\right\}\om=0.\NO
\eea
The solutions of these equations, which are non-singular as $X\to\infty$, are
respectively
\bea
e^i_j & = & e\sub{0}^i_jX^{-a}\left[(1+a)X+(1-a)X^{-1}\right]\\&&
\left\{1+\frac{\a}{8a^2}\left(a^2\left[(2a-1)X^2+(2a+1)X^{-2}\right]
-\frac{8(a^2+1)}{(a+1)X[(1+a)X+(1-a)X^{-1}]}\right.\right.\NO\\&&\left.\left.
-4\ln X\left[\frac{(2a^2-2a+1)(1+a)X+(2a^2+2a+1)(1-a)X^{-1}}
{(1+a)X+(1-a)X^{-1}}-(a^2-1)a\ln X\right]\right)\right.\NO\\
&&\left.+\co(\a^2)\rule[.2cm]{0cm}{.3cm}\right\},\NO\\\NO\\
\om & = & \om\sub{0}X^{-a}\left[(1+a)X-(1-a)X^{-1}\right]\\&&
\left\{1+\frac{\a}{8a^2}\left(a^2\left[(2a-3)X^2+(2a+3)X^{-2}\right]
-\frac{8[(a^2+1)-(a^2-1)(a^2-2)]}{(a+1)X[(1+a)X-(1-a)X^{-1}]}\right.\right.
\NO\\&&\left.\left.
-4\ln X\left[\frac{(2a^2-2a+1)(1+a)X-(2a^2+2a+1)(1-a)X^{-1}}
{(1+a)X-(1-a)X^{-1}}-(a^2-1)a\ln X\right]\right)\right.\NO\\
&&\left.+\co(\a^2)\rule[.2cm]{0cm}{.3cm}\right\},\NO
\eea
where $a=\sqrt{1+q^2}$ and $e\sub{0}^i_j$ and $\om\sub{0}$ are arbitrary
functions of $q$.

It is now straightforward to evaluate $E$ and $\Om$ using these exact
solutions. To isolate the desired coefficients $E\sub{3}$ and $\Om\sub{1}$ we
first need to determine the terms $E\sub{1},\;E\sub{2}$ as well as
$\Om\sub{0}$ and subtract them from the exact solutions for $E$ and $\Om$.
This can be done using the first order equations (\ref{firstorderequations})
and the dilatation operator as described in Section \ref{2pt-functions}.
One easily finds
\bea
&&E\sub{1}=\Om\sub{0}=0,\\
&&E\sub{2}=\frac{4}{l}(a^2-1)e^{-2A}=\frac{1}{4l}(a^2-1)(X-X^{-1})^2
\left[1-\frac{\a}{4}(X^2-X^{-2}+8\ln X)+\co(\a^2)\right].\NO
\eea
Subtracting these from the exact solutions for $E$ and $\Om$ we finally
determine
\bea\label{dynamic}
&&E\sub{3}=-\frac{1}{l}\left(\frac{\k}{\sqrt{2}}\right)^3\f_B^3a(a^2-1)
\left(1+\a\frac{(3a^2-1)}{2a^3(a^2-1)}+\co(\a^2)\right),\NO\\
&&\Om\sub{1}=\frac{1}{l}\frac{\k}{\sqrt{2}}\f_Ba^{-1}(a^2-1)
\left(1+\a\frac{(3a^2-1)(2a^2-1)}{2a^3(a^2-1)}+\co(\a^2)\right),
\eea
which allow one to evaluate the exact one-point functions with
linear sources (\ref{linearized_mom_half_ren}) and hence the two-point
functions.

\section{Multi-trace deformations in the large-$N$ limit and the 
AdS/CFT correspondence}
\label{multitrace}

Multi-trace operators in the AdS/CFT correspondence have been studied
extensively \cite{Witten:2001ua,Berkooz:2002ug,Muck:2002gm,
Minces:2002wp, Sever:2002fk, Gubser:2002vv, Aharony:2005sh, Elitzur:2005kz}.
Before we discuss such operators in the context of the 
AdS/CFT correspondence, however, it is useful to review some generic
field theoretic properties of multi-trace operators in quantum field theories
that admit a large-$N$ limit. These properties are independent of 
the AdS/CFT correspondence and will allow us to incorporate multi-trace
operators in the AdS/CFT correspondence in a very elegant way.  

We will first for completeness repeat the field theory argument given in
\cite{Elitzur:2005kz}, applicable to any quantum field theory that admits a
large-$N$ limit. Let $\co(x)$ be a local gauge-invariant single-trace
operator, with the trace taken in the adjoint for concreteness, normalized
such that $\langle \co\rangle =\co(N^0)$ as $N\to\infty$. The generating
functional of connected correlators, $W[J]$ is $\co(N^2)$ and so it is
convenient to write $W[J]=N^2w[J]$. In terms of the field theory action,
$S[\f]$, then
\be
e^{-N^2w[J]}=\int [d\f]e^{-S[\f]- N^2\int d^dxJ(x)\co(x)}.
\ee
Now define
\be
\s(x)\equiv \langle \co\rangle_J=\frac{\d w[J]}{\d J}.
\ee  
The effective action $\G[\s]=N^2\bar{\G}[\s]$ is given by
\be
e^{-N^2\bar{\G}[\s]}=\int [dJ]e^{-N^2w[J]+N^2\int d^dx J(x)\s(x)},
\ee
and
\be
J=-\frac{\d \bar{\G}[\s]}{\d \s}.
\ee

Consider now the deformed action $S_f[\f]=S[\f]+N^2\int d^dx f(\co)$. Then,
\bea
e^{-N^2w_f[J_f]} & = & \int [d\f]e^{-S[\f]- N^2\int d^dx(J_f\co+f(\co))}\NO\\
& = & \int [d\f]e^{-S[\f]- N^2\int d^dx(J\co+f(\co)-f'(\s)\co)}\NO\\
&\stackrel{N\to\infty}{\approx}&e^{-N^2w[J]}e^{- N^2\int d^dx(f(\s)
-\s f'(\s))},
\eea
where we introduced 
\be\label{sources}
J\equiv J_f+f'(\s),
\ee
in the second line in order to remove the linear term from $f(\co)$ so that
large-$N$ factorization can be used in the last step. It follows 
that in the large-$N$ limit, the generating functional of connected 
correlators in the deformed theory is given by
\bea
w_f[J_f] & = &\bar{\G}_f[\s]+\int d^dx J_f \s \NO\\
& = & w[J]+\int\left. d^dx \left(f(\s)-\s f'(\s)\right)\right|_{\s
=\d w[J]/\d J}.
\eea 
Moreover, 
\bea
e^{-N^2\bar{\G}_f[\s]} & = & \int [dJ_f]e^{-N^2 w_f[J_f]+N^2\int d^dx J_f \s}
\NO\\
& = & \int [dJ]e^{-N^2 w_f[J]}e^{-N^2\int d^dx (f(\s)-\s f'(\s))}
e^{N^2\int d^dx (J-f'(\s))\s}\NO\\
& = & e^{-N^2\bar{\G}[\s]-N^2\int d^dx f(\s)},
\eea
where we have used $[d J_f]=[dJ]$. Therefore,
\be
\bar{\G}_f[\s]=\bar{\G}[\s]+\int d^dxf(\s),
\ee
or equivalently
\be
V_{\rm eff}^f(\s)=V_{\rm eff}(\s)+N^2f(\s).
\ee

These results rely only on the existence of a large-$N$ limit and are
independent of the AdS/CFT correspondence. However, they allow for an 
elegant reformulation of Witten's \cite{Witten:2001ua} prescription
for incorporating multi-trace operators in the AdS/CFT correspondence.
Recall, that in the supergravity approximation, one computes the 
renormalized on-shell supergravity action $I[\f]$, which is a 
functional of $\f_-$ only since $\f_+$ is expressed in terms of $\f_-$
by requiring regularity in the interior. Equipped with the renormalized
on-shell supergravity action, the AdS/CFT prescription for multi-trace
operators can be stated as follows, depending on the $\D_\pm$ quantization: 

For the $\D=\D_+$ quantization, one identifies the generating functional 
of the undeformed theory as
\be
W[J]\equiv I[\f_-]|_{\f_-=J}.
\ee   
For the deformed theory, Witten's prescription amounts to setting
\bea\label{D+shift}
W_f[J_f] & \equiv & I[\f_-]|_{\f_-=\f_-^*(J_f)}\NO\\
& = & I[\f_-]|_{\f_-=J}+\int \left. d^dx \left(f(\s)-\s f'(\s)\right)
\right|_{\s=\d I[\f_-]/\d \f_-|_{\f_-=J}},
\eea  
where $\f_-^*(J_f)$ is the solution to
\be
\f_-=J_f+f'\left(\frac{\d I[\f_-]}{\d\f_-}\right).
\ee
Noting that $\f_-=J$ and $\frac{\d I[\f_-]}{\d\f_-}=\s$, this equation is
precisely equation (\ref{sources}), thus justifying this prescription
for the incorporation of multi-trace operators in the AdS/CFT correspondence.

For the $\D=\D_-$ quantization, one merely needs to replace $I[\f_-]$
with the Legendre transform $\bar{I}[\bar{\f}_-]$ defined in 
(\ref{legendre1},\ref{legendre2},\ref{legendre-constr}). Namely,
the generating functional of the undeformed theory is now given by
\be
\lbar{W}[\bar{J}]\equiv \bar{I}[\bar{\f}_-]|_{\bar{\f}_-=\bar{J}},
\ee
while the effective action is given by the on-shell action
\be
\lbar{\G}[\bar{\s}]=I[\bar{\s}].
\ee
For the deformed theory then
\be
\lbar{W}_{\bar{f}}[\bar{J}_{\bar{f}}]\equiv \bar{I}[\bar{\f}_-]|_{\bar{\f}_-
=\bar{\f}_-^*(\bar{J}_{\bar{f}})},
\ee
where $\bar{\f}_-^*(\bar{J}_{\bar{f}})$ is the solution to
\be
\bar{\f}_-=\bar{J}_{\bar{f}}+\bar{f}'\left(\frac{\d \bar{I}[\bar{\f}_-]}
{\d\bar{\f}_-}\right).
\ee
The effective action is then given by
\bea\label{D-shift}
\lbar{\G}_f[\bar{\s}]& = & \lbar{W}[\bar{J}_{\bar{f}}]-N^2\int d^dx 
\bar{J}_{\bar{f}}\bar{\s}\NO\\
& = & I[\bar{\s}]+N^2\int d^dx \bar{f}(\bar{\s}).
\eea

\end{document}